\documentclass[twocolumn,showkeys]{revtex4}
\usepackage[utf8]{inputenc}
\usepackage{graphicx}

\begin{document}

\title{Extended relativistic kinetic model composed of the scalar and two vector distribution functions: Application to the spin-electron-acoustic waves}

\author{Pavel A. Andreev}
\email{andreevpa@physics.msu.ru}
\affiliation{Department of General Physics, Faculty of physics, Lomonosov Moscow State University, Moscow, Russian Federation, 119991.}

\date{\today}

\begin{abstract}
Detailed deterministic derivation of the kinetic equations for the relativistic plasmas is given.
Focus is made on the dynamic of one-coordinate distribution functions of various tensor dimensions,
but the closed set of kinetic equations is constructed of three functions:
the scalar distribution function,
the vector distribution function of dipole moment,
and the vector distribution function of velocity (or the dipole moment in the momentum space).
All two-coordinate distributions functions are discussed as well.
They are presented together with their limits existing in the self-consistent field approximation.
The dynamics of the small amplitude spin-electron-acoustic waves in the dense degenerate plasmas is are studied within the kinetic model.
This work presents the deterministic approach to the derivation and interpretation of the kinetic equations.
So, no probability is introduced during the transition from the level of individual particles to the collective functions.
Problem of thermalization is not considered,
but we can see that the structure of kinetic equations is kept for the system before and after thermalization.
Hence, the kinetic equations can be used to approach this item.
\end{abstract}


\maketitle





\section{Introduction}

Kinetic theory is one of fundamental theories of the macroscopic phenomena in various physical systems
\cite{Klimontovich book},
\cite{Israel AP 79}, \cite{Romatschke IJMPE 10}, \cite{Kovtun JP A 12},
\cite{Denicol PRD 12}, \cite{Heinz ARNPS 13}, \cite{Florkowski RPP 18}, \cite{Denicol PRD 19}.
Particularly, is plays essential role in the plasma physics \cite{Hakim book Rel Stat Phys}.
Kinetic theory requires proper microscopic derivation,
where the evolution of the distribution function is traced from the evolution of individual particles.
The distribution function shows the relative positions of all particles in the six dimensional phase space
and it can be consistently defined in the deterministic way.
So, no probability theory is applied for the derivation or interpretation of the kinetic equations or the distribution function.

Such view on the kinetic theory is the extension of the classical hydrodynamics based on the tracing of the microscopic dynamics of individual particles
\cite{Drofa TMP 96}, \cite{Kuzmenkov CM 15}.

Classical mechanics gives us dynamics of particles in the 3N-dimensional configurational space.
The model of motion of particles is composed of the individual trajectories.
Each of them, formally, takes place in its own three dimensional space.
this picture of motion is not obvious
if we consider the Newtonian form of mechanics,
where we can use our own imagination to place all particles in the single three-dimensional space.
However, the Lagrangian and Hamiltonian forms of mechanics show the multidimensional structure of the theory more clearly.

Moreover, the construction of the mathematical model out of trajectories
(which are one dimensional mathematical objects)
distinguish the classical mechanics from the hydrodynamics or the electrodynamics,
which are the field theories.

The derivation of the hydrodynamics from the mechanics requires to represent the mechanics in the form of the field theory.
So, the mechanics would have the same mathematical structure as the electrodynamics,
which is essentially important for the dynamics of charged particles.

Refs. \cite{Drofa TMP 96} and \cite{Kuzmenkov CM 15} basically present the field form of the classical mechanics,
where the dynamics of particles takes form of the dynamics of material fields.
Inevitably, the field form of classical mechanics appears in the form of hydrodynamic equations.
These are equations of nonequilibrium hydrodynamics
which requires further reduction for the particular forms of the collective motion.

Particularly, it leads to the some generalized form of nonequilibrium thermodynamics.
And choosing regime of thermal equilibrium allows to get the first low of thermodynamics from the energy evolution equation.
The possibility to consider different regimes, like regimes close to the thermal equilibrium, and regimes out of equilibrium, shows
that this approach can be used for the study of the thermalization process in the physical systems
considering the dynamics of the systems in terms of the collective variables (material fields).
It would require proper truncation of the hydrodynamics out equilibrium considering higher rank material fields to include the mechanism of relaxation in the model.
However, it is an open problem
which can be addressed within this formulation.

Formulation of the mechanics in terms of the three-dimensional material fields leads to the hydrodynamics form of equations.
While we consider the relative positions and relative momentums of all particles
we go to the distribution of particles in the six dimensional space
(the phase space of coordinate and momentum).
This form of the classical mechanics is considered in this paper.

If we consider the microscopic evolution of point-like particles
we get the following distribution of the particle number in the coordinate space
\cite{Weinberg Gr 72}
\begin{equation}\label{RHD2022ClSCF concentration definition zero volume}
n_{m}(\textbf{r},t)=\sum_{i=1}^{N}\delta(\textbf{r}-\textbf{r}_{i}(t)), \end{equation}
where $\textbf{r}_{i}(t)$ is the coordinate of $i$-th particle.
This approach is suggested by Yu.L. Klimontovich \cite{Klimontovich book}, \cite{Weinberg Gr 72}.
While we use notation $n_{m}(\textbf{r},t)$ on the left-hand side of equation (\ref{RHD2022ClSCF concentration definition zero volume})
we should write $n_{m}(\textbf{r},\textbf{r}_{1}(t), ..., \textbf{r}_{N}(t))$.
So, the time dependence of function $n_{m}$ is directly constructed out of the time evolution of coordinates of particles.
The parametrization of the physical space is introduced here $\textbf{r}$.
So this form gives us an intermediate step from the mechanics of trajectories to the field form of classical mechanics.

Considering the microscopic distribution (\ref{RHD2022ClSCF concentration definition zero volume})
different authors applied some form of averaging (see for instance \cite{Klimontovich book}, \cite{deGrootFoundations1972}).
Sometimes, authors use some unspecified distributions $\langle n_{m}\rangle$
(see for instance \cite{Klimontovich book}).
Overwise authors use specific form of averaging via some distribution function
(see for instance \cite{deGrootFoundations1972}, \cite{Kuzelev UFN 99}).
Anyway, authors try to impose probabilistic interpretation of the procedure of averaging.
Let us mention
that the aim of averaging is to make the transition to the macroscopic scale
by averaging on the physically infinitesimal volume,
but usually it is being replaced by the averaging on some distribution function.

\begin{figure}\includegraphics[width=8cm,angle=0]{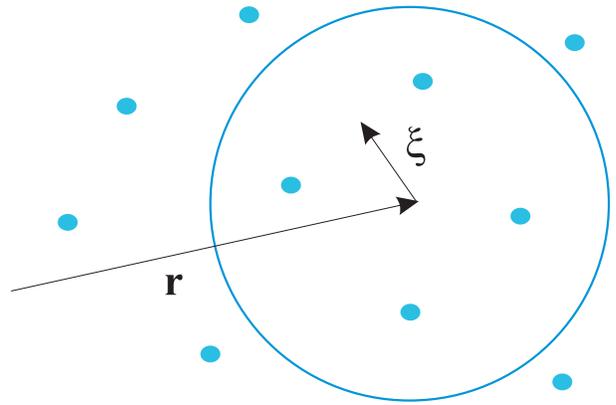}
\caption{\label{RHD2022ClSCF Fig 01}
The illustration of the $\Delta$-vicinity in the coordinate space.}
\end{figure}

Systematic, nonstatistical, definition of the averaging on the physically infinitesimal volume is suggested by L.S. Kuz'menkov \cite{Drofa TMP 96}
(see also \cite{Andreev PIERS 2012}, \cite{Andreev 2021 05})
\begin{equation}\label{RHD2022ClSCF concentration definition}
n_{e}(\textbf{r},t)=\frac{1}{\Delta}\int_{\Delta}d\mbox{\boldmath $\xi$}\sum_{i=1}^{N/2}\delta(\textbf{r}+\mbox{\boldmath $\xi$}-\textbf{r}_{i}(t)). \end{equation}
Equation (\ref{RHD2022ClSCF concentration definition}) contains vector $\mbox{\boldmath $\xi$}$
which scanning the physically infinitesimal volume
(it is illustrated in equation \ref{RHD2022ClSCF Fig 01}).
While the physically infinitesimal volume appears as the $\Delta$-vicinity of each point of space.
It is assumed that $N$ is the full number of particles in the system.
It is composed of two species: electrons with numbers $i\in [1,N/2]$ and protons $i\in [N/2+1,N]$.
We need to trace the evolution of each species in the system.
With no restrictions, we illustrate our derivation on the subsystem of electrons,
so subindex $e$ is dropped for all functions describing electrons in equations below $n_{e}\equiv n$.

Similar background is suggested for the kinetic theory.
So, the microscopic distribution function is composed of delta functions in the physical coordinate space and the momentum space
\begin{equation}\label{RHD2022ClSCF distr funct definition zero volume}
f_{m}(\textbf{r},\textbf{p},t)=\sum_{i=1}^{N}\delta(\textbf{r}-\textbf{r}_{i}(t))\delta(\textbf{p}-\textbf{p}_{i}(t)). \end{equation}
Transition to the macroscopic level is made similarly to equation (\ref{RHD2022ClSCF concentration definition}).
We introduce the physically infinitesimal volume in six-dimensional phase space
$$f(\textbf{r},\textbf{p},t)=
\frac{1}{\Delta}\frac{1}{\Delta_{p}}\times$$
\begin{equation}\label{RHD2022ClSCF distribution function definition}
\times\int_{\Delta,\Delta_{p}}
d\mbox{\boldmath $\xi$}
d\mbox{\boldmath $\eta$}
\sum_{i=1}^{N/2}
\delta(\textbf{r}+\mbox{\boldmath $\xi$}-\textbf{r}_{i}(t))
\delta(\textbf{p}+\mbox{\boldmath $\eta$}-\textbf{p}_{i}(t)), \end{equation}
where
$d\mbox{\boldmath $\eta$}$ is the element of volume in the momentum space,
with $\int_{\Delta_{p}}d\mbox{\boldmath $\eta$}$ integral over $\Delta_{p}$-vicinity in the momentum space,
$\Delta\equiv \Delta_{r}$ is the delta vicinity in the coordinate space.
Recent development of this approach can be found in Refs.
\cite{Andreev Ph 15}, \cite{Zubkov JPhCS 19}, \cite{Zubkov 2006}, \cite{Zubkov VNSU 22}.

This discussion is about the field form of the classic mechanics.
However, similar insight is imposed on the quantum mechanics.
So, the formulation of the many-particle quantum mechanics in the form of evolution of the material fields is developed
\cite{Maksimov QHM 99}, \cite{MaksimovTMP 2001}, \cite{Andreev LP 19}, \cite{Andreev LP 21 fermions},
\cite{Andreev LPL 21}, \cite{Andreev JPP 21}, \cite{Andreev Ch 21}, \cite{Andreev PoF 21}.

The discussion presented above is focused on the fundamental background of the hydrodynamics and kinetics,
since the theoretical part of this paper is based on the development of the kinetic theory.
However, the application of the kinetic theory to the high-density magnetized spin polarized degenerate electron gas is also considered.

Particularly, we focus on the spin-electron-acoustic waves (SEAWs).
Existence of the SEAWs in the partially spin-polarized degenerate plasmas was theoretically suggested in 2015
\cite{Andreev PRE 15}.
They appear as the longitudinal waves of density and electric field,
which are propagate parallel to the anisotropy direction in the magnetized electron gas
\cite{Andreev PRE 15}, \cite{Andreev PP 16 SSE kin}.

If we have no spin polarization
we find single longitudinal wave in the electron gas under assumption of the motionless ions.
It is the Langmuir wave,
which corresponds to the collective oscillations of electrons relatively motionless ions.
If we include the spin-polarization
we also find the SEAWs.
The physical picture of dynamics is following.
It corresponds to relative motion of electrons with different spin projections.
The spin polarization manifests itself in different partial concentrations of electrons with the particular spin polarization.
Hence, the relative oscillations of electrons with different amplitudes of concentration give
the oscillation of small proportion of electrons relatively motionless ions as well
\cite{Andreev PRE 15}, \cite{Andreev PP 16 SSE kin}.
The SEAWs show similarity to the spin-plasmons considered in two-dimensional structures
\cite{Ryan PRB 91}, \cite{Agarwal PRL 11}, \cite{Agarwal PRB 14},
\cite{Schober EPL 20}, \cite{Amico JPD 19}, \cite{Yang Ph 21}, \cite{Kreil CPP 18},
\cite{Kreil PRB 15}, \cite{Huang PRB 16}, \cite{Afanasiev PRB 22}.
Recently, the SEAWs have been considered in the degenerate electron gas
\cite{Andreev 2112},
while the background of applied model is given in Refs.
\cite{Andreev 2021 05}, \cite{Andreev 2021 09}, \cite{Andreev 2021 11}.

Increase of the concentration of electrons $n_{0e}$ up to values,
where the Fermi energy is proportional to the rest energy of electron
$\varepsilon_{Fe}=(3\pi^{2}n_{0e})^{2/3}\hbar^{2}/2m_{e}$$\sim m_{e}c^{2}$,
gives noticeable change in the dispersion dependence of the SEAWs.
It also corresponds to the change of the dispersion dependence of the Langmuir waves.
It shows the decrease of the relative frequency of the Langmuir wave and SEAW.
Moreover, the relativistic regime shows the growth of the amplitude of the concentration of the SEAWs relatively the amplitude of the Langmuir waves
for the chosen value of the electric field in these waves.


This paper is organized as follows.
In Sec. II the method of derivation of the kinetic equation from the microscopic motion is demonstrated
and the general structure of the kinetic equation is derived.
In Sec. III we present an approximate kinetic model,
where the contribution of physically infinitesimal volume is considered in minimal regime called the monopole regime.
In Sec. IV the selfconsistent field approximation of monopole regime is considered.
In Sec. V the multipole approximation in the relativistic kinetic equation is presented as the generalization of the monopole regime considered in Sec. III.
In Sec. VI the selfconsistent field approximation of the multipole approximation is presented.
In Sec. VII the kinetic equation for the evolution of the dipole moment vector distribution function is obtained.
In Sec. VIII the kinetic equation for the distribution function of velocity is derived.
In Sec. IX closed set of three kinetic equations is discussed.
In Sec. X the spin-electron-acoustic waves in the spin polarized electron gas of high density are considered.
In Sec. XI a brief summary of obtained results is presented.

\section{Derivation of the Vlasov equation tracing the microscopic motion of particles}

We need to consider the evolution of the distribution function.
It requires to present equation of motion of each particle in full details
\begin{equation}\label{RHD2022ClSCF Eq of Motion Newtor nonRel Gen}
\dot{\textbf{p}}_{i}(t)=\textbf{F}(\textbf{r}_{i}(t),t), \end{equation}
where
$\textbf{p}_{i}(t)=m_{i}\textbf{v}_{i}(t)/\sqrt{1-\textbf{v}_{i}^{2}(t)/c^{2}}$ is the relativistic momentum of $i$-th particle,
and $\textbf{F}(\textbf{r}_{i}(t),t)$ is the force acting on $i$-th particle from the electromagnetic field created by other particles in the system.
The equation of motion for each particle appears as the evolution of the momentum under action of the Lorentz force
\begin{equation}\label{RHD2022ClSCF Eq of Motion Newton Rel}
\textbf{F}(\textbf{r}_{i}(t),t)=
\biggl(q_{i}\textbf{E}(\textbf{r}_{i}(t),t)
+\frac{1}{c}q_{i}[\textbf{v}_{i}(t), \textbf{B}(\textbf{r}_{i}(t),t)]
\biggr), \end{equation}
where
$\textbf{E}_{i}=\textbf{E}_{i,ext}+\textbf{E}_{i,int}$,
$\textbf{B}_{i}=\textbf{B}_{i,ext}+\textbf{B}_{i,int}$,
$\textbf{E}_{i}=\textbf{E}(\textbf{r}_{i}(t),t)$,
and
$\textbf{B}_{i}=\textbf{B}(\textbf{r}_{i}(t),t)$,
while the external fields are included along with the field of interaction of particles.
The electric $\textbf{E}_{i,int}$ and magnetic $\textbf{B}_{i,int}$ fields caused by particles surrounding the $i$-th particle
are $\textbf{E}_{i,int}=-\nabla_{i}\varphi(\textbf{r}_{i}(t),t)-\frac{1}{c}\partial_{t}\textbf{A}(\textbf{r}_{i}(t),t)$
and $\textbf{B}_{i,int}=\nabla_{i}\times \textbf{A}(\textbf{r}_{i}(t),t)$
with \cite{Landau v2}
\begin{equation}\label{RHD2022ClSCF varphi via Green function rel}
\varphi(\textbf{r}_{i}(t),t)=\sum_{j\neq i}q_{j}\int \frac{\delta(t-t'-\frac{1}{c}\mid \textbf{r}_{i}(t)-\textbf{r}_{j}(t')\mid)}{\mid \textbf{r}_{i}(t)-\textbf{r}_{j}(t')\mid}dt',
\end{equation}
and
\begin{equation}\label{RHD2022ClSCF A via Green function rel}
\textbf{A}(\textbf{r}_{i}(t),t)=\sum_{j\neq i}q_{j}\int \frac{\delta(t-t'-\frac{1}{c}\mid \textbf{r}_{i}(t)-\textbf{r}_{j}(t')\mid)}{\mid \textbf{r}_{i}(t)-\textbf{r}_{j}(t')\mid}
\frac{\textbf{v}_{j}(t')}{c}dt'.
\end{equation}
Equations (\ref{RHD2022ClSCF varphi via Green function rel})
and (\ref{RHD2022ClSCF A via Green function rel})
allow to introduce
the Green function of the retarding electromagnetic interaction \cite{Landau v2}
\begin{equation}\label{RHD2022ClSCF Green function rel}
\tilde{G}_{ij}=
\frac{\delta(t-t'-\frac{1}{c}\mid \textbf{r}_{i}(t)-\textbf{r}_{j}(t')\mid)}{\mid \textbf{r}_{i}(t)-\textbf{r}_{j}(t')\mid}. \end{equation}

We consider the high-temperature plasmas ,
where electrons (and may be other species) have relativistic temperatures
comparable with the rest energy of the electron $m_{e}c^{2}$ (of the particle of corresponding species).
Obtaining such huge temperatures of the species would lead to high degree of ionization of the atomic objects.
Moreover, the two-particle interactions (collision-like processes) of the high-energy electrons
with the ions would change the electron configurations of ions or provide additional ionization.
These dynamical processes should contribute in the model.
To avoid these complexifications
we consider the hydrogen plasmas,
where we have electrons and protons only.

To find the equation for evolution of the distribution function (\ref{RHD2022ClSCF distribution function definition})
we consider the time derivative of this function to obtain the following equation
\begin{widetext}
$$\partial_{t}f(\textbf{r},\textbf{p},t)+
\nabla\cdot\frac{1}{\Delta}\frac{1}{\Delta_{p}}
\int_{\Delta,\Delta_{p}}
d\mbox{\boldmath $\xi$}
d\mbox{\boldmath $\eta$}
\sum_{i=1}^{N/2} \dot{\textbf{r}}_{i}(t)
\delta(\textbf{r}+\mbox{\boldmath $\xi$}-\textbf{r}_{i}(t))
\delta(\textbf{p}+\mbox{\boldmath $\eta$}-\textbf{p}_{i}(t))$$
\begin{equation}\label{RHD2022ClSCF kin eq delta I}
+\nabla_{\textbf{p}}\cdot\frac{1}{\Delta}\frac{1}{\Delta_{p}}
\int_{\Delta,\Delta_{p}}
d\mbox{\boldmath $\xi$}
d\mbox{\boldmath $\eta$}
\sum_{i=1}^{N/2} \dot{\textbf{p}}_{i}(t)
\delta(\textbf{r}+\mbox{\boldmath $\xi$}-\textbf{r}_{i}(t))
\delta(\textbf{p}+\mbox{\boldmath $\eta$}-\textbf{p}_{i}(t))=0. \end{equation}
\end{widetext}
The second (third) term in this equation appears
as the result of action of the time derivative on the delta-function containing the coordinate (the momentum).

The second term in equation (\ref{RHD2022ClSCF kin eq delta I}) contains the following function
\begin{equation}\label{RHD2022ClSCF function F def REL}
\langle\textbf{v}_{i}(t)\rangle\equiv\frac{1}{\Delta}\frac{1}{\Delta_{p}}
\int_{\Delta,\Delta_{p}}
d\mbox{\boldmath $\xi$}
d\mbox{\boldmath $\eta$}
\sum_{i=1}^{N/2} \textbf{v}_{i}(t)
\delta_{\textbf{r}i}
\delta_{\textbf{p}i},
\end{equation}
where
$\textbf{v}_{i}(t)=d \textbf{r}_{i}(t)/dt$,
$\delta_{\textbf{r}i}\equiv\delta(\textbf{r}+\mbox{\boldmath $\xi$}-\textbf{r}_{i}(t))$, and
$\delta_{\textbf{p}i}\equiv\delta(\textbf{p}+\mbox{\boldmath $\eta$}-\textbf{p}_{i}(t))$.
For the further representation of this function
we apply relativistic expression for the velocity of particle via its momentum
$\textbf{v}_{i}(t)=\textbf{p}_{i}(t)c/\sqrt{\textbf{p}_{i}^{2}(t)+m_{i}^{2}c^{2}}$,
where additional replacement of the momentum $\textbf{p}_{i}(t)$ on $\textbf{p}+\mbox{\boldmath $\eta$}$ can be applied.
It gives the representation of function $\langle\textbf{v}_{i}(t)\rangle$:
\begin{equation}\label{RHD2022ClSCF function F repr 1 REL}
\langle\textbf{v}_{i}(t)\rangle=\frac{1}{\Delta}\frac{1}{\Delta_{p}}
\int_{\Delta,\Delta_{p}}
d\mbox{\boldmath $\xi$}
d\mbox{\boldmath $\eta$}
\sum_{i=1}^{N/2} \frac{(\textbf{p}+\mbox{\boldmath $\eta$})c}{\sqrt{(\textbf{p}+\mbox{\boldmath $\eta$})^{2}+m_{i}^{2}c^{2}}}
\delta_{\textbf{r}i}
\delta_{\textbf{p}i}.
\end{equation}
Here, our goal is the extraction of traditional for the physical kinetics terms $\textbf{v}\cdot f$.
Hence, expression (\ref{RHD2022ClSCF function F repr 1 REL}) is represented in required form
\begin{equation}\label{RHD2022ClSCF function F repr 2 REL}
\langle\textbf{v}_{i}(t)\rangle =\textbf{v}\cdot f(\textbf{r},\textbf{p},t)+\textbf{F}(\textbf{r},\textbf{p},t), \end{equation}
where
\begin{equation}\label{RHD2022ClSCF function F repr 2b REL}\textbf{F}(\textbf{r},\textbf{p},t)\equiv \langle \Delta\textbf{v}_{i}(t)\rangle\end{equation}
with
$$\langle \Delta\textbf{v}_{i}\rangle\equiv\frac{1}{\Delta}\frac{1}{\Delta_{p}}
\int_{\Delta,\Delta_{p}}
d\mbox{\boldmath $\xi$}
d\mbox{\boldmath $\eta$}
\sum_{i=1}^{N/2}
\biggl(\frac{\mbox{\boldmath $\eta$}c}{\sqrt{\textbf{p}^{2}+m_{i}^{2}c^{2}}}$$
\begin{equation}\label{RHD2022ClSCF function F repr 2b REL dop}
-\frac{\textbf{p}(\textbf{p}\cdot\mbox{\boldmath $\eta$})c}{(\sqrt{\textbf{p}^{2}+m_{i}^{2}c^{2}})^{3}}\biggr)
\delta_{\textbf{r}i}
\delta_{\textbf{p}i},
\end{equation}
with
$\textbf{p}=m_{s}\textbf{v}/\sqrt{1-\textbf{v}^{2}/c^{2}}$ and
$\textbf{v}=\textbf{p}c/\sqrt{\textbf{p}^{2}+m_{s}^{2}c^{2}}$.


We include representation (\ref{RHD2022ClSCF function F repr 2 REL}) in equation (\ref{RHD2022ClSCF kin eq delta I}).
We also include equations of motion of the individual particles
(\ref{RHD2022ClSCF Eq of Motion Newtor nonRel Gen})-(\ref{RHD2022ClSCF A via Green function rel})
in the last term in equation (\ref{RHD2022ClSCF kin eq delta I}):
\begin{widetext}
$$\partial_{t}f(\textbf{r},\textbf{p},t)+
(\textbf{v}\cdot\nabla)f(\textbf{r},\textbf{p},t)
+\nabla\cdot \textbf{F}(\textbf{r},\textbf{p},t)
+\frac{q_{s}}{m_{s}}\frac{1}{\Delta}\frac{1}{\Delta_{p}}
\int_{\Delta,\Delta_{p}}
d\mbox{\boldmath $\xi$}
d\mbox{\boldmath $\eta$}
\sum_{i=1}^{N/2} 
\biggl[\textbf{E}_{ext}(\textbf{r}+\mbox{\boldmath $\xi$},t)$$
\begin{equation}\label{RHD2022ClSCF kin eq delta II}
+\frac{1}{c}[(\textbf{v}+\Delta\textbf{v}_{i})\times \textbf{B}_{ext}(\textbf{r}+\mbox{\boldmath $\xi$},t)]
+q_{s'}\sum_{j=1,j\neq i}^{N}\int d t'\biggl(-\nabla_{\textbf{r}}-\frac{1}{c}\frac{\textbf{v}_{j}(t')}{c}
\partial_{t}+\frac{1}{c^{2}}[\textbf{v}_{i}(t)\times [\nabla_{\textbf{r}}\times\textbf{v}_{j}(t')]]\biggr)
G(\textbf{r}+\mbox{\boldmath $\xi$}-\textbf{r}_{j}(t))\biggr]
\delta_{\textbf{r}i}
\cdot\nabla_{\textbf{p}}\delta_{\textbf{p}i}=0, \end{equation}
where
we use symbol $\hat{\partial_{t}}$ instead of the time derivative due to the change of the structure of argument of the Green function $G$.
We replaced $\textbf{r}_{i}(t)$ by $\textbf{r}+\mbox{\boldmath $\xi$}$
in the arguments of the external electric field, the external magnetic field, and the Green function using the delta-function $\delta_{\textbf{r}i}$.
Initially the action of the time derivative has the following form
\begin{equation}\label{RHD2022ClSCF}
\partial_{t}G(t,t',\textbf{r}_{i}(t),\textbf{r}_{j}(t'))
=\frac{\delta'}{\mid \textbf{r}_{i}(t)-\textbf{r}_{j}(t')\mid},
\end{equation}
where $\delta'$ is the derivative of the delta function on its argument.
It is also can be rewritten in the following form
\begin{equation}\label{RHD2022ClSCF}
\partial_{t}G(t,t',\textbf{r},\textbf{r}')
=\frac{\delta'}{\mid \textbf{r}+\mbox{\boldmath $\xi$}-\textbf{r}'-\mbox{\boldmath $\xi$}'\mid},
\end{equation}
For the introduction of $\mbox{\boldmath $\xi$}'$ see the equations below.
%

Next, we use representation (\ref{RHD2022ClSCF function F repr 2 REL}) in terms describing the interaction:
$$\partial_{t}f+\textbf{v}\cdot\nabla f+\nabla\cdot \textbf{F}
+\frac{q_{s}}{m_{s}}\frac{1}{\Delta}\frac{1}{\Delta_{p}}
\int_{\Delta,\Delta_{p}}
d\mbox{\boldmath $\xi$}
d\mbox{\boldmath $\eta$}
\sum_{i=1}^{N/2} \biggl(\textbf{E}(\textbf{r}+\mbox{\boldmath $\xi$},t)
+\frac{1}{c}[(\textbf{v}+\Delta\textbf{v}_{i})\times \textbf{B}(\textbf{r}+\mbox{\boldmath $\xi$},t)]\biggr)
\delta_{\textbf{r}i}
\cdot\nabla_{\textbf{p}}\delta_{\textbf{p}i}$$
$$+\frac{q_{s}}{m_{s}}q_{s'}
\int
d\textbf{r}' d\textbf{p}'
\int_{\Delta,\Delta_{p}}
\frac{d\mbox{\boldmath $\xi$}
d\mbox{\boldmath $\eta$}
d\mbox{\boldmath $\xi$}'
d\mbox{\boldmath $\eta$}'}{\Delta^{2}\Delta_{p}^{2}}
\sum_{i=1}^{N/2}\sum_{j=1, j\neq i}^{N}\int d t'
\biggl(-\nabla_{\textbf{r}}-\frac{1}{c}\frac{(\textbf{v}'+\Delta\textbf{v}_{j}')}{c}
\partial_{t}$$
\begin{equation}\label{RHD2022ClSCF kin eq delta III}
-\frac{1}{c^{2}}[(\textbf{v}+\Delta\textbf{v}_{i})\times [(\textbf{v}'+\Delta\textbf{v}_{j}')\times\nabla_{\textbf{r}}]]\biggr)
G(\textbf{r}+\mbox{\boldmath $\xi$}-\textbf{r}'-\mbox{\boldmath $\xi$}')
\delta_{\textbf{r}i}
\cdot\nabla_{\textbf{p}} \delta_{\textbf{p}i}
\delta_{\textbf{r}'j}
\delta_{\textbf{p}'j}
=0, \end{equation}
\end{widetext}
where
$\delta_{\textbf{r}'j}\equiv\delta(\textbf{r}'+\mbox{\boldmath $\xi$}'-\textbf{r}_{j}(t))$, and
$\delta_{\textbf{p}'j}\equiv\delta(\textbf{p}'+\mbox{\boldmath $\eta$}'-\textbf{p}_{j}(t))$.

Presented kinetic equation (\ref{RHD2022ClSCF kin eq delta III}) contains the deviation of coordinates $\mbox{\boldmath $\xi$}$
from the center of the $\Delta$-vicinity $\textbf{r}$.
It also contain the deviation of the velocity $\Delta\textbf{v}_{i}$ from the value corresponding to the center of the $\Delta_{\textbf{p}}$-vicinity
in the momentum space $\textbf{v}=\textbf{p}c/\sqrt{\textbf{p}^{2}+m_{s}^{2}c^{2}}$.
This is the intermediate general representation of the kinetic equation
which is considered below under some additional assumptions.


\section{Monopole approximation of the kinetic equation for the scalar distribution function}

To consider the monopole approximation of the kinetic equation
we need to neglect contribution of $\mbox{\boldmath $\xi$}$, $\mbox{\boldmath $\xi$}'$,
$\mbox{\boldmath $\eta$}$ (including $\Delta\textbf{v}_{i}$),
and $\mbox{\boldmath $\eta$}'$ (including $\Delta\textbf{v}_{j}'$) in dynamical functions.
However, on this step, we present the kinetic equation,
where the monopole approximation is considered for the space variables only.
Hence, we neglect contribution of $\mbox{\boldmath $\xi$}$, and $\mbox{\boldmath $\xi$}'$,
but we keep the contribution of $\mbox{\boldmath $\eta$}$:
$$\partial_{t}f+\textbf{v}\cdot\nabla f+\nabla\cdot \textbf{F}
+\frac{q_{s}}{m_{s}}\biggl(\textbf{E}(\textbf{r},t)
+\frac{1}{c}[\textbf{v}\times \textbf{B}(\textbf{r},t)]\biggr)\cdot\nabla_{\textbf{p}}f$$
$$+\frac{q_{s}}{m_{s}c}(\nabla_{\textbf{p}}\cdot[\textbf{F}(\textbf{r},\textbf{p},t)\times\textbf{B}(\textbf{r},t)])$$
$$+\frac{q_{s}}{m_{s}}q_{s'}
\int
d\textbf{r}' d\textbf{p}'
\int_{\Delta,\Delta_{p}}
\frac{d\mbox{\boldmath $\xi$}
d\mbox{\boldmath $\eta$}
d\mbox{\boldmath $\xi$}'
d\mbox{\boldmath $\eta$}'}{\Delta^{2}\Delta_{p}^{2}}
\sum_{i=1}^{N/2}\sum_{j=1, j\neq i}^{N}\int d t'
\biggl(-\nabla_{\textbf{r}}$$
\begin{equation}\label{RHD2022ClSCF kin eq monopole r1}
-\frac{\textbf{v}'}{c^{2}} \partial_{t}
+\frac{1}{c^{2}}[\textbf{v}\times [\nabla\times\textbf{v}']]\biggr)
G(\textbf{r},\textbf{r}')
\delta_{\textbf{r}i}
\cdot\nabla_{\textbf{p}} \delta_{\textbf{p}i}
\delta_{\textbf{r}'j}
\delta_{\textbf{p}'j}=0. \end{equation}
Transition from equation (\ref{RHD2022ClSCF kin eq delta III}) to equation (\ref{RHD2022ClSCF kin eq monopole r1}) includes the transformation like
$\textbf{E}(\textbf{r}+\mbox{\boldmath $\xi$},t)\approx \textbf{E}(\textbf{r},t)$
and
\begin{equation}\label{RHD2022ClSCF}
\frac{1}{\Delta}\frac{1}{\Delta_{p}}
\int_{\Delta,\Delta_{p}}
d\mbox{\boldmath $\xi$}
d\mbox{\boldmath $\eta$}
\sum_{i=1}^{N/2} \textbf{E}(\textbf{r},t)
\delta_{\textbf{r}i}
\cdot\nabla_{\textbf{p}}\delta_{\textbf{p}i}
=\textbf{E}(\textbf{r},t) \cdot\nabla_{\textbf{p}}f,
\end{equation}
where the electric field $\textbf{E}(\textbf{r},t)$ is placed outside of the integral,
while the rest is the derivative of the distribution function on the momentum.

Equation (\ref{RHD2022ClSCF kin eq monopole r1}) allows to introduce the two-particle (or in other terms two-coordinate) distribution functions.
To the best of my knowledge, majority of papers and books, including my own works, use notion "two-particle distribution functions".
It can be confusing to some extend
since the model describes the many-particle systems.
Hence, the distribution function and the two-particle distribution function actually describe the dynamics of whole system.
Sometimes, the two-particle distribution functions are called "two-coordinate distribution functions".
It looks more logical
since it contains two coordinates $\textbf{r}$ and $\textbf{r}'$
(or more exactly two coordinates in the phase space $\{\textbf{r},\textbf{p}\}$ and $\{\textbf{r}', \textbf{p}'\}$).

Let us present the following kinetic equation in the monopole approximation both on coordinate and momentum
(we do it due to the technical reasons,
since we do not want to introduce several two-coordinate distribution functions,
but we show the complete picture below):
$$\partial_{t}f+\textbf{v}\cdot\nabla f+\nabla\cdot \textbf{F}
+\frac{q_{s}}{m_{s}}\biggl(\textbf{E}
+\frac{1}{c}[\textbf{v}\times \textbf{B}]\biggr)\cdot\nabla_{\textbf{p}}f$$
$$+\frac{q_{s}}{m_{s}c}(\nabla_{\textbf{p}}\cdot[\textbf{F}\times\textbf{B}])
+\frac{q_{s}}{m_{s}}q_{s'}
\int d\textbf{r}' d\textbf{p}'
\int d t' \biggl(-\nabla_{\textbf{r}}$$
\begin{equation}\label{RHD2022ClSCF kin eq monopole r2}
-\frac{\textbf{v}'}{c^{2}}\partial_{t}
+\frac{1}{c^{2}}[\textbf{v}\times [\nabla\times\textbf{v}']]\biggr)
G(\textbf{r},\textbf{r}')\cdot\nabla_{\textbf{p}}f_{2}
=0, \end{equation}
where
$q_{s'}f_{2}=q_{e}f_{2,ee}+q_{i}f_{2,ei}$
and
$$f_{2,ee}(\textbf{r},\textbf{r}',\textbf{p},\textbf{p}',t,t')=
\int_{\Delta,\Delta_{p}}
\frac{d\mbox{\boldmath $\xi$}
d\mbox{\boldmath $\eta$}
d\mbox{\boldmath $\xi$}'
d\mbox{\boldmath $\eta$}'}{\Delta^{2}\Delta_{p}^{2}}
\times$$
\begin{equation}\label{RHD2022ClSCF}
\times
\sum_{i=1}^{N/2}\sum_{j=1, j\neq i}^{N/2}
\delta_{\textbf{r}i}
\delta_{\textbf{p}i}
\delta_{\textbf{r}'j}
\delta_{\textbf{p}'j}\equiv \langle\langle1\rangle\rangle \end{equation}
is the two-coordinate distribution function.
The compact notation is also introduced for the two-coordinate distribution function
constructed of double brackets $\langle\langle1\rangle\rangle$.
We use this notation below for other two-coordinate distribution functions.


\section{The selfconsistent field approximation in classic monopole kinetics for the equation of evolution of scalar distribution function}

In order to consider the selfconsistent field (the mean-field) approximation
we need to split the two-coordinate distribution function $f_{2,ee}(\textbf{r},\textbf{r}',\textbf{p},\textbf{p}',t,t')$
into the product of two one-coordinate distribution functions
$f_{2,ee}(\textbf{r},\textbf{r}',\textbf{p},\textbf{p}',t,t')=f(\textbf{r},\textbf{p},t)\cdot f(\textbf{r}',\textbf{p}',t')$.
Hence, the term containing interaction
in equation
(\ref{RHD2022ClSCF kin eq monopole r2})
reappears in the following form
$$\frac{q_{s}}{m_{s}}q_{s'}\nabla_{\textbf{p}}f(\textbf{r},\textbf{p},t)\cdot\biggl(
-\nabla_{\textbf{r}}\int d\textbf{r}' d\textbf{p}'\int d t' G(\textbf{r},\textbf{r}')f(\textbf{r}',\textbf{p}',t')$$
$$-\frac{1}{c^{2}}\int d\textbf{r}' d\textbf{p}'\int d t'\textbf{v}'
\partial_{t}G(\textbf{r},\textbf{r}')f(\textbf{r}',\textbf{p}',t')$$
\begin{equation}\label{RHD2022ClSCF kin eq monopole r2 SCF}
+\frac{1}{c^{2}}[\textbf{v}\times [\nabla\times \int d\textbf{r}' d\textbf{p}'\int d t'\textbf{v}']]
G(\textbf{r},\textbf{r}')f(\textbf{r}',\textbf{p}',t')=0. \end{equation}
It allows to introduce the scalar and vector potentials of the self-consistent electromagnetic field
\begin{equation}\label{RHD2022ClSCF}  \varphi_{int}(\textbf{r},t)=\int d\textbf{r}' d\textbf{p}'\int d t' G(\textbf{r},\textbf{r}') f(\textbf{r}',\textbf{p}',t')\end{equation}
and
\begin{equation}\label{RHD2022ClSCF} \textbf{A}_{int}(\textbf{r},t)=\frac{1}{c}\int d\textbf{r}' d\textbf{p}'\int d t'\textbf{v}'
G(\textbf{r},\textbf{r}') f(\textbf{r}',\textbf{p}',t').
\end{equation}
Some discussion on the selfconsistent field approximation for the kinetics based on the  deterministic microscopic motion of particles is gives in Ref. \cite{Andreev 2204}
for the nonrelativistic kinetics and hydrodynamics.

So, we find the well-known form of the Vlasov kinetic equation for the system of relativistic particles \cite{Vlasov JETP 38}
\begin{equation}\label{RHD2022ClSCF Vlasov eq Coulomb}
\partial_{t}f+\textbf{v}\cdot\nabla f
+q_{s}\biggl(\textbf{E}+\frac{1}{c}\textbf{v}\times\textbf{B}\biggr)
\cdot\frac{\partial f}{\partial \textbf{p}}=0
,\end{equation}
where
$\textbf{E}=\textbf{E}_{ext}+\textbf{E}_{int}$,
$\textbf{B}=\textbf{B}_{ext}+\textbf{B}_{int}$,
the electric field is caused by the distribution of charges in the coordinate space:
$\textbf{E}_{int}=-\nabla\varphi-\partial_{t}\textbf{A}_{int}/c$,
$\textbf{B}_{int}=\nabla \times \textbf{A}_{int}$.
Equation (\ref{RHD2022ClSCF Vlasov eq Coulomb}) appears to be coupled with the electromagnetic field:
$\nabla\times \textbf{E}_{int}=-\partial_{t}\textbf{B}_{int}/c$,
$\nabla\cdot \textbf{B}_{int}=0$,
\begin{equation}\label{RHD2021ClLM rot B kin monopole}
\nabla\times \textbf{B}_{int}= \partial_{t}\textbf{E}_{int}/c +(4\pi/c) \sum_{s}^{} q_{s}\int \textbf{v}f_{s}(\textbf{r},\textbf{p},t)d\textbf{p},
\end{equation}
and
\begin{equation}\label{RHD2021ClLM div E kin monopole}
\nabla\cdot \textbf{E}_{int}=4\pi \sum_{s}^{} q_{s}\int f_{s}(\textbf{r},\textbf{p},t)d\textbf{p}.\end{equation}

\section{Multipole approximation in the relativistic kinetic equation for the scalar distribution function
(the Vlasov equation)}


\subsection{The interaction of particles with the external electromagnetic field}

The multipole expansion of the electric field $\textbf{E}_{ext}(\textbf{r}+\mbox{\boldmath $\xi$},t)$,
the magnetic field $\textbf{B}_{ext}(\textbf{r}+\mbox{\boldmath $\xi$},t)$,
and the Green function $G(\textbf{r}-\textbf{r}'+\mbox{\boldmath $\xi$}-\mbox{\boldmath $\xi$}',t-t')$
leads to appearance of new distribution functions of different tensor ranks.
Let us introduce these functions before
we find they appearance from the expansion of equation (\ref{RHD2022ClSCF kin eq delta III}) on $\mbox{\boldmath $\xi$}$ and $\mbox{\boldmath $\xi$}'$
up to the second order on these vectors.

Firstly, we introduce the vector distribution function of the electric dipole moment
(divided by the charge $q_{s}$)
or the displacement of particles relatively the center of the $\Delta$-vicinity:
\begin{equation}\label{RHD2022ClSCF d rp def}
d^{a}(\textbf{r},\textbf{p},t)=\langle\xi^{a}\rangle=
\int_{\Delta,\Delta_{p}}
\frac{d\mbox{\boldmath $\xi$}
d\mbox{\boldmath $\eta$}}{\Delta\Delta_{p}}
\sum_{i=1}^{N} \xi^{a}
\delta_{\textbf{r}i}
\delta_{\textbf{p}i}.
\end{equation}

We also include in our analysis the distribution function of the electric quadrupole moment
(divided by the charge $q_{s}$)
\begin{equation}\label{RHD2022ClSCF Q rp def}
Q^{ab}(\textbf{r},\textbf{p},t)=\langle\xi^{a}\xi^{b}\rangle=
\int_{\Delta,\Delta_{p}}
\frac{d\mbox{\boldmath $\xi$}
d\mbox{\boldmath $\eta$}}{\Delta\Delta_{p}}
\sum_{i=1}^{N}
\xi^{a}\xi^{b}
\delta_{\textbf{r}i}
\delta_{\textbf{p}i}.
\end{equation}

Presented functions contain vector $\mbox{\boldmath $\xi$}$ scanning the $\Delta$-vicinity.
Functions (\ref{RHD2022ClSCF d rp def}) and (\ref{RHD2022ClSCF Q rp def}) appear
as two examples of the infinite set of distribution functions containing different degrees of vector $\mbox{\boldmath $\xi$}$
(the product of different number of projections to construct an element of tensor of corresponding rank).

As it is demonstrated by equations (\ref{RHD2022ClSCF function F def REL})-(\ref{RHD2022ClSCF function F repr 2b REL})
we can find the velocity of particle under integral defining the distribution function.
We can also find the product of several projections of the velocity
(of the same particle, in order to get one-coordinate distribution function)
or the product of the projection of the velocity on the projections of vector $\mbox{\boldmath $\xi$}$.

One of such distribution functions appearing in our derivation is the distribution function of flux of the electric dipole moment
(divided by the charge $q_{s}$)
$$J_{D}^{ab}(\textbf{r},\textbf{p},t)=\langle v_{i}^{a}(t)\xi^{b}\rangle=\frac{1}{\Delta^{2}}\frac{1}{\Delta_{p}^{2}}\times$$
\begin{equation}\label{RHD2022ClSCF J D rp def}
\times\int_{\Delta,\Delta_{p}}
d\mbox{\boldmath $\xi$}
d\mbox{\boldmath $\eta$}
d\mbox{\boldmath $\xi$}'
d\mbox{\boldmath $\eta$}'
\sum_{i,j=1, j\neq i}^{N} v_{i}^{a}(t) \xi^{b}
\delta_{\textbf{r}i}
\delta_{\textbf{p}i}
\delta_{\textbf{r}'j}
\delta_{\textbf{p}'j}.
\end{equation}
It is more useful to extract the velocity corresponding to the center of $\Delta_{\textbf{p}}$-vicinity in the momentum space,
like we make it for $\langle v_{i}^{a}(t)\rangle$ in equations (\ref{RHD2022ClSCF function F def REL})-(\ref{RHD2022ClSCF function F repr 2b REL}).
Therefore, function (\ref{RHD2022ClSCF J D rp def}) leads to the following distribution function:
$$j_{D}^{ab}(\textbf{r},\textbf{p},t)=\langle \Delta v_{i}^{a}(t)\xi^{b}\rangle=\frac{1}{\Delta^{2}}\frac{1}{\Delta_{p}^{2}}\times$$
\begin{equation}\label{RHD2022ClSCF j D rp def}
\times\int_{\Delta,\Delta_{p}}
d\mbox{\boldmath $\xi$}
d\mbox{\boldmath $\eta$}
d\mbox{\boldmath $\xi$}'
d\mbox{\boldmath $\eta$}'
\sum_{i,j=1, j\neq i}^{N} (\Delta v_{i}^{a}(t)) \xi^{b}
\delta_{\textbf{r}i}
\delta_{\textbf{p}i}
\delta_{\textbf{r}'j}
\delta_{\textbf{p}'j}.
\end{equation}

Let us introduce one more distribution function, which is the third rank tensor.
It is the distribution function of flux of the electric quadrupole moment
(divided by the charge $q_{s}$)
$$J_{Q}^{abc}(\textbf{r},\textbf{p},t)=\langle v_{i}^{a}(t)\xi^{b}\xi^{c}\rangle=\frac{1}{\Delta^{2}}\frac{1}{\Delta_{p}^{2}}\times$$
\begin{equation}\label{RHD2022ClSCF J Q rp def}
\times\int_{\Delta,\Delta_{p}}
d\mbox{\boldmath $\xi$}
d\mbox{\boldmath $\eta$}
d\mbox{\boldmath $\xi$}'
d\mbox{\boldmath $\eta$}'
\sum_{i,j=1, j\neq i}^{N} v_{i}^{a}(t) \xi^{b} \xi^{c}
\delta_{\textbf{r}i}
\delta_{\textbf{p}i}
\delta_{\textbf{r}'j}
\delta_{\textbf{p}'j}.
\end{equation}
For this distribution function,
we can also introduce
the reduced flux of the electric quadrupole moment on the velocities in the local frame $\Delta v_{i}^{a}(t)$:
$j_{Q}^{abc}(\textbf{r},\textbf{p},t)=\langle \Delta v_{i}^{a}(t)\xi^{b}\xi^{c}\rangle$.

Functions (\ref{RHD2022ClSCF d rp def})-(\ref{RHD2022ClSCF J Q rp def}) appear in the Vlasov-like kinetic equation
for the scalar distribution function.
Firstly, these functions appear at the analysis of the action of the external electromagnetic field on the system of charged particles.
Here, we need to consider the fourth term in equation (\ref{RHD2022ClSCF kin eq delta III}).
We apply the expansion of the external electromagnetic field on the deviations of coordinate $\mbox{\boldmath $\xi$}$
from the center of the $\Delta$-vicinity of point $\textbf{r}$.
This expansion requires slow change of the fields $\textbf{E}$ and $\textbf{B}$ over the diameter $\sqrt[3]{\Delta}$ of the $\Delta$-vicinity.
Particularly, if we consider the electromagnetic wave
we assume
that the wavelength $\lambda$ is larger than the diameter $\sqrt[3]{\Delta}$: $\lambda\geq\sqrt[3]{\Delta}$.

For instance, if we consider diameter $\sqrt[3]{\Delta}\approx0.1$ $\mu$m
we get estimation on the minimal value of the frequency of the electromagnetic wave of order of $\omega_{max}\sim10^{16}$ s$^{-1}$.
So, we can consider electromagnetic waves with frequencies up to the frequencies of the visible light,
but the ultraviolet radiation is our of the range of applicability of the model.
If we deal with the dense medium
we can choose the delta-vicinity of the smaller value.
So, the range of the ultraviolet radiation can be included in our analysis.

During derivation of the kinetic equations
we consider the expansion up to the second order on $\xi^{a}$:
$\textbf{E}(\textbf{r}+\mbox{\boldmath $\xi$},t)
=\textbf{E}(\textbf{r},t)$$+\xi_{a}\partial^{a}_{\textbf{r}}\textbf{E}(\textbf{r},t)$
$+(1/2)\xi_{a}\xi_{b}\partial^{a}_{\textbf{r}}\partial^{b}_{\textbf{r}}\textbf{E}(\textbf{r},t)+...$,
and
$\textbf{B}(\textbf{r}+\mbox{\boldmath $\xi$},t)
=\textbf{B}(\textbf{r},t)$$+\xi_{a}\partial^{a}_{\textbf{r}}\textbf{B}(\textbf{r},t)$
$+(1/2)\xi_{a}\xi_{b}\partial^{a}_{\textbf{r}}\partial^{b}_{\textbf{r}}\textbf{B}(\textbf{r},t)+...$.
Hence, the fourth term in equation (\ref{RHD2022ClSCF kin eq delta III}) can be rewritten as
$$\Omega_{0}=\frac{q_{s}}{m_{s}}
\langle (\textbf{E}_{ext}(\textbf{r}+\mbox{\boldmath $\xi$},t)$$
\begin{equation}\label{RHD2021ClLM Omaga 0 a}
+\frac{1}{c}[\textbf{v}\times\textbf{B}_{ext}(\textbf{r}+\mbox{\boldmath $\xi$},t)]
+\frac{1}{c}[\Delta\textbf{v}_{i}\times\textbf{B}_{ext}(\textbf{r}+\mbox{\boldmath $\xi$},t)])\nabla_{\textbf{p}}\rangle \end{equation}
before the expansion.
After described expansion we obtain
$$\Omega_{0}=\frac{q_{s}}{m_{s}}\biggl\{\textbf{E}_{ext}(\textbf{r},t)\nabla_{\textbf{p}}f(\textbf{r},\textbf{p},t)
+(\partial^{a}_{\textbf{r}}\textbf{E}_{ext})\nabla_{\textbf{p}}d^{a}(\textbf{r},\textbf{p},t)$$
$$+\frac{1}{2}(\partial^{a}_{\textbf{r}}\partial^{b}_{\textbf{r}}\textbf{E}_{ext})\nabla_{\textbf{p}}Q^{ab}(\textbf{r},\textbf{p},t)
+\frac{1}{c}[\textbf{v}\times\textbf{B}_{ext}(\textbf{r},t)]\nabla_{\textbf{p}}f(\textbf{r},\textbf{p},t)$$
$$+\frac{1}{c}[\textbf{v}\times(\partial^{a}_{\textbf{r}}\textbf{B}_{ext})]\nabla_{\textbf{p}}d^{a}(\textbf{r},\textbf{p},t)$$
$$+\frac{1}{c}[\textbf{v}\times(\partial^{a}_{\textbf{r}}\partial^{b}_{\textbf{r}}\textbf{B}_{ext})]
\nabla_{\textbf{p}}Q^{ab}(\textbf{r},\textbf{p},t)$$
$$+\frac{1}{c}\varepsilon^{abc}\biggl[B^{b}_{ext}\nabla_{\textbf{p}}^{c}F^{a}(\textbf{r},\textbf{p},t)
+(\partial^{d}_{\textbf{r}}B^{b}_{ext})\nabla_{\textbf{p}}^{c}j^{ad}(\textbf{r},\textbf{p},t)$$
\begin{equation}\label{RHD2021ClLM Omaga 0 b}
+(\partial^{d}_{\textbf{r}}\partial^{f}_{\textbf{r}}B^{b}_{ext})
\nabla_{\textbf{p}}^{c}j^{adf}(\textbf{r},\textbf{p},t)\biggr]\biggr\}. \end{equation}

\subsection{The interparticle interaction}

Let us introduce function $\Omega$
which presents a term in the kinetic equation describing the interparticle interaction
(the last term in equation (\ref{RHD2022ClSCF kin eq delta III})).
We split this term on three parts
$\Omega=\Omega_{1}+\Omega_{2}+\Omega_{3}$
which are
$\Omega_{1}=-(q_{s}/m_{s})\langle    \nabla_{i}\varphi_{i,int}\cdot\nabla_{p,i}\rangle$,
$\Omega_{2}=-(q_{s}/m_{s}c)\langle  \partial_{t}\textbf{A}_{i,int} \cdot\nabla_{p,i}\rangle$,
$\Omega_{3}=(q_{s}/m_{s}c) \varepsilon^{abc} \varepsilon^{cfg}\langle  v_{i}^{b}(t)\partial_{i}^{f}A_{i,int}^{g} \cdot\nabla_{p,i}\rangle$,
where we use short notation for the average on the physically infinitesimal volume, like
\begin{equation}\label{RHD2021ClLM}
\langle \nabla_{i}\varphi_{i,int}\cdot\nabla_{p,i}\rangle
=\int_{\Delta_{\textbf{r}}\Delta_{\textbf{p}}}
\frac{d\mbox{\boldmath $\xi$}d\mbox{\boldmath $\eta$}}{\Delta_{\textbf{r}}\Delta_{\textbf{p}}}
\sum_{i}\nabla_{i}\varphi_{i,int}\delta_{\textbf{r}i}\cdot\nabla_{p,i}\delta_{\textbf{p}i}. \end{equation}
Here, functions $\varphi_{i,int}$ and $\textbf{A}_{i,int}$ are given by
equations (\ref{RHD2022ClSCF varphi via Green function rel}) and (\ref{RHD2022ClSCF A via Green function rel}).

Let us consider the multipole expansion for each of function $\Omega_{j}$ (j=1,2,3).
We start this presentation with $\Omega_{1}$
which can be represented in the following form
$$\Omega_{1}=
-\int \frac{d\textbf{r}' d\textbf{p}'}{(\Delta_{\textbf{r}}\Delta_{\textbf{p}})^{2}}
\int dt'
\int_{\Delta_{\textbf{r}}\Delta_{\textbf{p}}}
d\mbox{\boldmath $\xi$}d\mbox{\boldmath $\eta$}
\int_{\Delta_{\textbf{r}'}\Delta_{\textbf{p}'}}
d\mbox{\boldmath $\xi$}'d\mbox{\boldmath $\eta$}'\times$$
\begin{equation}\label{RHD2021ClLM Omega 1}
\times\sum_{i,j,j\neq i} \frac{q_{i}q_{j}}{m_{s}}
\nabla_{\textbf{r}}\biggl(
G(t,t',\textbf{r}+\mbox{\boldmath $\xi$}, \textbf{r}'+\mbox{\boldmath $\xi$}')\biggr)
\delta_{\textbf{r}i}\cdot\nabla_{p,i}\delta_{\textbf{p}i}
\delta_{\textbf{r}'j}\delta_{\textbf{p}'j}.
\end{equation}

We make expansion of the Green functions on
$\mbox{\boldmath $\xi$}$ and $\mbox{\boldmath $\xi$}'$.
We also expand the velocity of particle $v_{i}^{b}$ on the deviation from the average velocity $\mbox{\boldmath $\eta$}$.
For the Green function $G(t,t',\textbf{r}+\mbox{\boldmath $\xi$}, \textbf{r}'+\mbox{\boldmath $\xi$}')$
we have $G=G_{0}+\xi^{b}G_{1,b}+(1/2)\xi^{b}\xi^{c}G_{2,bc}$,
where
$G_{0}=\delta(t-t'-\mid \textbf{r}-\textbf{r}'\mid/c)/\mid \textbf{r}-\textbf{r}'\mid$,
$G_{1,b}=\partial_{\textbf{r}}^{b}G_{0}$,
and
$G_{2,bc}=\partial_{\textbf{r}}^{b}\partial_{\textbf{r}}^{c}G_{0}$.

Function $\Omega_{1}$ can be represented in terms of two-coordinate distribution functions:
$$\Omega_{1}=-\frac{q_{s}q_{s'}}{m_{s}}\nabla_{\textbf{p}}
\int d\textbf{r}' d\textbf{p}'
\int dt'
\biggl[(\nabla_{\textbf{r}}G_{0})f_{2,ss'}(\textbf{r},\textbf{r}',\textbf{p},\textbf{p}',t,t')$$
$$+(\nabla_{\textbf{r}}G_{1}^{\beta})
\biggl(d_{2,ss'}^{b}(\textbf{r},\textbf{r}',\textbf{p},\textbf{p}',t,t')
-d_{2,ss'}^{b}(\textbf{r}',\textbf{r},\textbf{p},\textbf{p}',t,t')\biggr)$$
$$+\frac{1}{2}(\nabla_{\textbf{r}}G_{2}^{bc})
\biggl(q_{2,ss'}^{bc}(\textbf{r},\textbf{r}',\textbf{p},\textbf{p}',t,t')
+q_{2,ss'}^{bc}(\textbf{r}',\textbf{r},\textbf{p},\textbf{p}',t,t')$$
\begin{equation}\label{RHD2021ClLM Omega 1 fin}
-D_{2,ss'}^{bc}(\textbf{r},\textbf{r}',\textbf{p},\textbf{p}',t,t')
-D_{2,ss'}^{bc}(\textbf{r}',\textbf{r},\textbf{p},\textbf{p}',t,t')\biggr)\biggr],
\end{equation}
where we use the following notations for the two-coordinate distribution functions
\begin{equation}\label{RHD2022ClSCF}
d_{2,ss'}^{b}(\textbf{r},\textbf{r}',\textbf{p},\textbf{p}',t,t')\equiv \langle\langle \xi^{b}\rangle\rangle
\rightarrow d^{b}_{s}(\textbf{r},\textbf{p},t) \cdot f_{s'}(\textbf{r}',\textbf{p}',t'), \end{equation}
\begin{equation}\label{RHD2022ClSCF}
q_{2,ss'}^{bc}(\textbf{r},\textbf{r}',\textbf{p},\textbf{p}',t,t')\equiv \langle\langle \xi^{b}\xi^{c}\rangle\rangle
\rightarrow q_{s}^{bc}(\textbf{r},\textbf{p},t)\cdot f_{s'}(\textbf{r}',\textbf{p}',t'), \end{equation}
and
\begin{equation}\label{RHD2022ClSCF}
D_{2,ss'}^{bc}(\textbf{r},\textbf{r}',\textbf{p},\textbf{p}',t,t')\equiv \langle\langle \xi^{b}\xi'^{c}\rangle\rangle
\rightarrow d^{b}_{s}(\textbf{r},\textbf{p},t)\cdot d_{s'}^{c}(\textbf{r}',\textbf{p}',t'). \end{equation}


Potential part of the electric force $q_{s}\textbf{E}$ of the interparticle interaction is presented
in terms of the two-coordinate distribution function.
Next, we make same representation for the part of the electric force $q_{s}\textbf{E}$ expressed via the vector potential.
Therefore, let us demonstrate the explicit form of $\Omega_{2}$:
$$\Omega_{2}= -\int \frac{d\textbf{r}' d\textbf{p}'}{(\Delta_{\textbf{r}}\Delta_{\textbf{p}})^{2}}
\int dt'
\int_{\Delta_{\textbf{r}}\Delta_{\textbf{p}}}
d\mbox{\boldmath $\xi$}d\mbox{\boldmath $\eta$}
\int_{\Delta_{\textbf{r}'}\Delta_{\textbf{p}'}}
d\mbox{\boldmath $\xi$}'d\mbox{\boldmath $\eta$}'\times$$
\begin{equation}\label{RHD2021ClLM Omega 2}
\times\sum_{i,j,j\neq i} \frac{q_{i}q_{j}}{m_{s}c^{2}}
\partial_{t}\biggl(\textbf{v}_{j}(t')
G(t,t',\textbf{r}+\mbox{\boldmath $\xi$}, \textbf{r}'+\mbox{\boldmath $\xi$}')\biggr)
\delta_{\textbf{r}i}\cdot\nabla_{p,i}\delta_{\textbf{p}i}
\delta_{\textbf{r}'j}\delta_{\textbf{p}'j}.
\end{equation}

Function $\Omega_{2}$ can be approximately represented in terms of two-coordinate distribution functions
after expansion on $\mbox{\boldmath $\xi$}$ and $\mbox{\boldmath $\xi$}'$:
$$\Omega_{2}=-\frac{1}{c^{2}}\frac{q_{s}q_{s'}}{m_{s}}
\partial_{\textbf{p}}^{a}
\int d\textbf{r}' d\textbf{p}'
\int dt'
\biggl[(\partial_{t}G_{0})
\langle\langle v_{i}^{a}(t)\rangle\rangle$$
$$+(\partial_{t}G_{1}^{b})
\biggl(\langle\langle \xi^{b}v_{j}^{a}(t')\rangle\rangle
-\langle\langle \xi'^{b}v_{j}^{a}(t')\rangle\rangle\biggr)
+\frac{1}{2}(\partial_{t}G_{2}^{bc})\times$$
\begin{equation}\label{RHD2021ClLM Omega 2 fin}
\times\biggl(\langle\langle \xi^{b}\xi^{c}v_{j}^{a}(t')\rangle\rangle
+\langle\langle \xi'^{b}\xi'^{c}v_{j}^{a}(t') \rangle\rangle
-2\langle\langle \xi^{b}\xi'^{c}v_{j}^{a}(t')\rangle\rangle
\biggr)\biggr],
\end{equation}
where we use the following notations for the two-coordinate distribution functions
\begin{equation}\label{RHD2022ClSCF}\langle\langle v_{i}^{a}(t)\rangle\rangle
\rightarrow j^{a}_{s}(\textbf{r},\textbf{p},t)\cdot f_{s'}(\textbf{r}',\textbf{p}',t'), \end{equation}
\begin{equation}\label{RHD2022ClSCF}\langle\langle v_{j}^{a}(t')\rangle\rangle
\rightarrow f_{s}(\textbf{r},\textbf{p},t)\cdot j_{s'}^{a}(\textbf{r}',\textbf{p}',t'), \end{equation}
\begin{equation}\label{RHD2022ClSCF} \langle\langle \xi^{b}v_{i}^{a}(t)\rangle\rangle
\rightarrow J_{D,s}^{ab}(\textbf{r},\textbf{p},t)\cdot f_{s'}(\textbf{r}',\textbf{p}',t'), \end{equation}
\begin{equation}\label{RHD2022ClSCF} \langle\langle \xi^{b}v_{j}^{a}(t')\rangle\rangle
\rightarrow d_{s}^{b}(\textbf{r},\textbf{p},t)\cdot j^{a}_{s'}(\textbf{r}',\textbf{p}',t'), \end{equation}
\begin{equation}\label{RHD2022ClSCF} \langle\langle \xi'^{b}v_{j}^{a}(t')\rangle\rangle
\rightarrow f_{s}(\textbf{r},\textbf{p},t)\cdot J_{D,s'}^{ab}(\textbf{r}',\textbf{p}',t'), \end{equation}
\begin{equation}\label{RHD2022ClSCF} \langle\langle \xi^{b}\xi^{c}v_{j}^{a}(t')\rangle\rangle
\rightarrow q^{bc}_{s}(\textbf{r},\textbf{p},t)\cdot j_{s'}^{a}(\textbf{r}',\textbf{p}',t'), \end{equation}
\begin{equation}\label{RHD2022ClSCF} \langle\langle \xi'^{b}\xi'^{c}v_{j}^{a}(t') \rangle\rangle
\rightarrow f_{s}(\textbf{r},\textbf{p},t)\cdot J_{Q,s'}^{abc}(\textbf{r}',\textbf{p}',t'), \end{equation}
and
\begin{equation}\label{RHD2022ClSCF} \langle\langle \xi^{b}\xi'^{c}v_{j}^{a}(t')\rangle\rangle
\rightarrow d_{s}^{b}(\textbf{r},\textbf{p},t)\cdot J_{D,s'}^{ac}(\textbf{r}',\textbf{p}',t'). \end{equation}


Finally, we present the explicit form of the magnetic part of the interparticle interaction:
$$\Omega_{3}= \frac{1}{c^{2}} \varepsilon^{abc}\varepsilon^{cfg}
\frac{1}{(\Delta_{\textbf{r}}\Delta_{\textbf{p}})^{2}}
\int d\textbf{r}' d\textbf{p}'
\int dt'$$
$$\int_{\Delta_{\textbf{r}}\Delta_{\textbf{p}}}
d\mbox{\boldmath $\xi$}d\mbox{\boldmath $\eta$}
\int_{\Delta_{\textbf{r}'}\Delta_{\textbf{p}'}}
d\mbox{\boldmath $\xi$}'d\mbox{\boldmath $\eta$}'
\sum_{i,j,j\neq i} \frac{q_{i}q_{j}}{m_{s}}
v_{i}^{b}(t)v_{j}^{g}(t')
\times$$
\begin{equation}\label{RHD2021ClLM Omega 3}
\times
\partial_{\textbf{r}}^{f}\biggl(
G(\textbf{r}+\mbox{\boldmath $\xi$}, \textbf{r}'+\mbox{\boldmath $\xi$}')\biggr)
\delta_{\textbf{r}i}\cdot\partial_{p,i}^{a}\delta_{\textbf{p}i}
\delta_{\textbf{r}'j}\delta_{\textbf{p}'j}
\end{equation}


Here, we make expansion on $\mbox{\boldmath $\xi$}$ and $\mbox{\boldmath $\xi$}'$
and make interpretation of found terms via the two-coordinate distribution functions:
$$\Omega_{3}=
\frac{1}{c^{2}}\frac{q_{s}q_{s'}}{m_{s}}
\varepsilon^{abc}\varepsilon^{cfg}
\partial_{\textbf{p}}^{a}
\int d\textbf{r}' d\textbf{p}'
\int dt'
\biggl[(\partial_{\textbf{r}}^{f} G_{0})
\langle\langle v_{i}^{b}(t)v_{j}^{g}(t')\rangle\rangle$$
$$+(\partial_{\textbf{r}}^{f}G_{1}^{k})
\biggl(\langle\langle \xi^{k}v_{i}^{b}(t)v_{j}^{g}(t')\rangle\rangle
-\langle\langle \xi'^{k}v_{i}^{b}(t)v_{j}^{g}(t')\rangle\rangle\biggr)$$
$$+\frac{1}{2}(\partial_{\textbf{r}}^{f}G_{2}^{kl})
\biggl(\langle\langle \xi^{k}\xi^{l}v_{i}^{b}(t)v_{j}^{g}(t')\rangle\rangle$$
\begin{equation}\label{RHD2021ClLM Omega 3 fin}
+\langle\langle \xi'^{k}\xi'^{l}v_{i}^{b}(t)v_{j}^{g}(t')\rangle\rangle
-2\langle\langle \xi^{k}\xi'^{l}v_{i}^{b}(t)v_{j}^{g}(t')\rangle\rangle \biggr)\biggr],
\end{equation}
where we use no specific notations for the two-coordinate distribution functions
since we have two many two-coordinate distribution functions.
We have no reason to use special symbol for each of them.
We also have following limits for the self-consistent field approximation:
\begin{equation}\label{RHD2022ClSCF} \langle\langle v_{i}^{b}(t)v_{j}^{g}(t')\rangle\rangle
\rightarrow j^{b}(\textbf{r},\textbf{p},t)\cdot j^{g}(\textbf{r}',\textbf{p}',t'), \end{equation}
\begin{equation}\label{RHD2022ClSCF} \langle\langle \xi^{a}v_{i}^{b}(t)v_{j}^{g}(t')\rangle\rangle
\rightarrow J_{D}^{ba}(\textbf{r},\textbf{p},t)\cdot j^{g}(\textbf{r}',\textbf{p}',t'), \end{equation}
\begin{equation}\label{RHD2022ClSCF} \langle\langle \xi'^{a}v_{i}^{b}(t)v_{j}^{g}(t')\rangle\rangle
\rightarrow j^{b}(\textbf{r},\textbf{p},t)\cdot J_{D}^{ga}(\textbf{r}',\textbf{p}',t'), \end{equation}
\begin{equation}\label{RHD2022ClSCF} \langle\langle \xi^{s}\xi^{l}v_{i}^{b}(t)v_{j}^{g}(t')\rangle\rangle
\rightarrow J^{bsl}_{Q}(\textbf{r},\textbf{p},t)\cdot j^{g}(\textbf{r}',\textbf{p}',t'), \end{equation}
\begin{equation}\label{RHD2022ClSCF} \langle\langle \xi'^{s}\xi'^{l}v_{i}^{b}(t)v_{j}^{g}(t')\rangle\rangle
\rightarrow j^{b}(\textbf{r},\textbf{p},t)\cdot J^{gsl}_{Q}(\textbf{r}',\textbf{p}',t'), \end{equation}
and
\begin{equation}\label{RHD2022ClSCF} \langle\langle \xi^{s}\xi'^{l}v_{i}^{b}(t)v_{j}^{g}(t')\rangle\rangle
\rightarrow J^{bs}_{D}(\textbf{r},\textbf{p},t)\cdot J_{D}^{gl}(\textbf{r}',\textbf{p}',t'). \end{equation}


To conclude this section, we point out
that we made the generalization of equation (\ref{RHD2022ClSCF kin eq monopole r2}) up to the account of the qudrupole moment distribution function
in the multipole expansion
while equation (\ref{RHD2022ClSCF kin eq monopole r2}) includes the distribution of charger (and currents as well) in the monopole approximation.
Presented generalization is demonstrated by four terms labeled as $\Omega_{0}$, $\Omega_{1}$, $\Omega_{2}$, and $\Omega_{3}$.

\section{Selfconsistent field approximation in the multipole approximation for the scalar distribution function evolution equation}

We consider the dynamics of charged  particles.
Therefore, the mean-field or the self-consistent field approximation gives the major contribution in the dynamics of system.
Hence, we consider all two-coordinate distribution functions as the product of the corresponding one-coordinate distribution functions.
Therefore, we combine equations (\ref{RHD2021ClLM Omaga 0 b}),
(\ref{RHD2021ClLM Omega 1 fin}), (\ref{RHD2021ClLM Omega 2 fin}), (\ref{RHD2021ClLM Omega 3 fin})
represented in the selfconsistent field approximation
and find the following equation for the scalar distribution function:
$$\partial_{t}f_{s}+
\textbf{v}\cdot\nabla f_{s}
+\nabla\cdot \textbf{F}_{s}
+q_{s}\biggl(\textbf{E}+\frac{1}{c}\textbf{v}\times\textbf{B}\biggr)\cdot\frac{\partial f_{s}}{\partial \textbf{p}}$$
$$+q_{s}\biggl(\partial^{b}\textbf{E}+\frac{1}{c}\textbf{v}\times\partial^{b}\textbf{B}\biggr)
\cdot\frac{\partial d_{s}^{b}}{\partial \textbf{p}}
+q_{s}\biggl(\partial^{b}\partial^{c}\textbf{E}$$
$$
+\frac{q_{s}}{c}\textbf{v}\times\partial^{b}\partial^{c}\textbf{B}\biggr)\cdot\frac{\partial Q_{s}}{\partial \textbf{p}}
+\frac{q_{s}}{c}\varepsilon^{abc}\biggl(B^{c}\cdot\partial_{a,\textbf{p}}
F_{s}^{b}  $$
\begin{equation}\label{RHD2022ClSCF Vlasov eq multipole SCF appr}
+\partial^{d}B^{c}\cdot \partial_{a,\textbf{p}}  j_{D,s}^{bd}
+\partial^{d}\partial^{f}B^{c}\cdot \partial_{a,\textbf{p}}  j_{Q,s}^{bdf}  \biggr)
=0,\end{equation}
where
the velocity is extracted from functions $j_{s}^{a}$, $j_{D,s}^{ab}$, and $j_{Q,s}^{abc}$.
We also have
$\textbf{E}=\textbf{E}_{ext}+\textbf{E}_{int}$,
and
$\textbf{B}=\textbf{B}_{ext}+\textbf{B}_{int}$.

The integral terms presenting the sources of the electromagnetic field are written as $\textbf{E}_{int}$ and $\textbf{B}_{int}$,
while these functions obey the Maxwell equations in accordance with the explicitly found integral expressions for the electromagnetic field
$\{\textbf{E}_{int},\textbf{B}_{int}\}$
via the one-coordinate distribution functions:
\begin{equation}\label{RHD2021ClLM Maxwell div B and rot E}
\begin{array}{cc}
  \nabla\cdot\textbf{B}_{int}=0, & \nabla\times \textbf{E}_{int}=-\frac{1}{c}\partial_{t}\textbf{B}_{int},
\end{array} \end{equation}
$$(\nabla\times \textbf{B}_{int})^{a}=\frac{1}{c}\partial_{t}E_{int} ^{a}
+\frac{4\pi}{c}\sum_{s}^{} q_{s}\biggl(
\int
j_{s}^{a}(\textbf{r},\textbf{p},t)
d\textbf{p}$$
\begin{equation}\label{RHD2021ClLM Maxwell rot B kin}
+\partial^{b}\int
J_{s,D}^{ab}(\textbf{r},\textbf{p},t)
d\textbf{p}
+\frac{1}{2} \partial^{b}\partial^{c}\int
J_{s,Q}^{abc}(\textbf{r},\textbf{p},t)
d\textbf{p}
+...\biggr),
\end{equation}
$$\nabla\cdot \textbf{E}_{int}=4\pi \sum_{s}^{} q_{s}\biggl(\int f_{s}(\textbf{r},\textbf{p},t)d\textbf{p}$$
\begin{equation}\label{RHD2021ClLM div E kin}
+\partial^{b}\int d_{s}^{b}(\textbf{r},\textbf{p},t)d\textbf{p}
+\frac{1}{2} \partial^{b}\partial^{c}\int Q_{s}^{bc}(\textbf{r},\textbf{p},t)d\textbf{p}
+...\biggr). \end{equation}
The selfconsistent field approximation shows
that all introduced distribution functions appear as the sources of the electromagnetic field in the Maxwell equations.

\section{Equation for the evolution of the dipole moment distribution function}

Let us repeat the definition of the distribution function of dipole
moment
\begin{equation}\label{RHD2022ClSCF d rp def REPEATED}
d^{a}(\textbf{r},\textbf{p},t)=\frac{1}{\Delta}\frac{1}{\Delta_{p}}
\int_{\Delta,\Delta_{p}}
d\mbox{\boldmath $\xi$}
d\mbox{\boldmath $\eta$}
\sum_{i=1}^{N} \xi^{a}
\delta_{\textbf{r}i}
\delta_{\textbf{p}i}
\end{equation}
If we want to consider the evolution of the distribution function of dipole moment
we can consider the time derivative of function $\langle \xi^{a}\rangle$ (\ref{RHD2022ClSCF d rp def REPEATED}).
We can consider alternative function $\langle r^{a}(t)\rangle$$=r^{a} f(\textbf{r},\textbf{p},t)+ \langle \xi^{a}\rangle$.

The time derivative of function (\ref{RHD2022ClSCF d rp def REPEATED}) has the following structure
\begin{equation}\label{RHD2021ClLM dipole moment time derivative}
\partial_{t}d^{a}
+\partial^{b}_{\textbf{r}}\langle \xi^{a}v_{i,b}(t)\rangle
+\partial^{b}_{\textbf{p}}\langle \xi^{a}\dot{p}_{i,b}(t)\rangle
=0.\end{equation}

The second term in this kinetic equation can be represented
via functions (\ref{RHD2022ClSCF J D rp def}) or (\ref{RHD2022ClSCF j D rp def}).
While the last term in this equation requires longer discussion similar
to the analysis of interaction in the Vlasov equation presented above.

Substituting the time derivative of the momentum $\dot{p}_{i}^{b}$ into equation (\ref{RHD2021ClLM dipole moment time derivative})
(equation of motion of each particle is given by equation
(\ref{RHD2022ClSCF Eq of Motion Newtor nonRel Gen}))-(\ref{RHD2022ClSCF A via Green function rel})
we obtain
\begin{widetext}
$$\partial_{t}d^{a}
+\partial^{b}_{\textbf{r}} J_{D}^{ba}
+q_{s}[\langle \xi^{a}E^{b}_{ext}(\textbf{r}+\mbox{\boldmath $\xi$},t)\partial_{b,\textbf{p}}\rangle
+\frac{1}{c}\varepsilon^{bcd}\langle \xi^{a}v_{i}^{c}(t)B^{d}_{ext}(\textbf{r}+\mbox{\boldmath $\xi$},t)\partial_{b,\textbf{p}}\rangle]$$
$$-q_{s}q_{s'}\int d\textbf{r}' d\textbf{p}'
\int dt'
\langle\langle \xi^{a} (\partial_{\textbf{r}}^{b}G(t,t',\textbf{r}+\mbox{\boldmath $\xi$},\textbf{r}'+\mbox{\boldmath $\xi$}')) \partial_{b,\textbf{p}}\rangle\rangle
-\frac{q_{s}q_{s'}}{c^{2}}\int d\textbf{r}' d\textbf{p}'
\int dt'
\langle\langle \xi^{a}v_{j}^{b}(t') (\partial_{t}G(t,t',\textbf{r}+\mbox{\boldmath $\xi$},\textbf{r}'+\mbox{\boldmath $\xi$}')) \partial_{b,\textbf{p}}\rangle\rangle$$
\begin{equation}\label{RHD2021ClLM}
+\frac{q_{s}q_{s'}}{c^{2}}
\varepsilon^{bcd}\varepsilon^{dfg}
\int d\textbf{r}' d\textbf{p}'
\int dt'
\langle\langle \xi^{a}v_{i}^{c}(t)v_{j}^{g}(t') (\partial_{\textbf{r}}^{f}G(t,t',\textbf{r}+\mbox{\boldmath $\xi$},\textbf{r}'+\mbox{\boldmath $\xi$}')) \partial_{b,\textbf{p}}\rangle\rangle
=0.\end{equation}


Next, we make expansion on $\mbox{\boldmath $\xi$}$ and $\mbox{\boldmath $\xi$}'$
in order to get the multipole expansion.
We keep terms up to the electric quadrupole moment and find:
$$\partial_{t}d^{a}
+\partial^{b}_{\textbf{r}} J_{D}^{ba}
+q_{s}\biggl[E^{b}_{ext}(\textbf{r},t)\partial_{b,\textbf{p}}\langle \xi^{a}\rangle
+ (\partial^{c}_{\textbf{r}}E^{b}_{ext}(\textbf{r},t))\partial_{b,\textbf{p}}\langle \xi^{a}\xi^{c}\rangle
+\frac{1}{c}\varepsilon^{bcd}B^{d}_{ext}(\textbf{r},t)\partial_{b,\textbf{p}}\langle \xi^{a}v_{i}^{c}(t)\rangle$$
$$+\frac{1}{c}\varepsilon^{bcd}(\partial^{f}_{\textbf{r}}B^{d}_{ext}(\textbf{r},t))\partial_{b,\textbf{p}}\langle \xi^{a}\xi^{f}v_{i}^{c}(t)\rangle\biggr]
-q_{s}q_{s'}\int d\textbf{r}' d\textbf{p}'
\int dt'
(\partial_{\textbf{r}}^{b}G(t,t',\textbf{r},\textbf{r}')) \partial_{b,\textbf{p}}
\langle\langle \xi^{a} \rangle\rangle$$
$$-q_{s}q_{s'}\int d\textbf{r}' d\textbf{p}'
\int dt'
(\partial_{\textbf{r}}^{b}\partial_{\textbf{r}}^{c}G(t,t',\textbf{r},\textbf{r}')) \partial_{b,\textbf{p}}
\langle\langle \xi^{a}\xi^{c} \rangle\rangle
+q_{s}q_{s'}\int d\textbf{r}' d\textbf{p}'
\int dt'
(\partial_{\textbf{r}}^{b}\partial_{\textbf{r}}^{c}G(t,t',\textbf{r},\textbf{r}')) \partial_{b,\textbf{p}}
\langle\langle \xi^{a}\xi'^{c} \rangle\rangle$$
$$-\frac{q_{s}q_{s'}}{c^{2}}\int d\textbf{r}' d\textbf{p}'
\int dt'
(\partial_{t}G(t,t',\textbf{r},\textbf{r}')) \partial_{b,\textbf{p}}
\langle\langle \xi^{a}v_{j}^{b}(t') \rangle\rangle
-\frac{q_{s}q_{s'}}{c^{2}}\int d\textbf{r}' d\textbf{p}'
\int dt'
(\partial_{t}\partial_{\textbf{r}}^{c}G(t,t',\textbf{r},\textbf{r}')) \partial_{b,\textbf{p}}
\langle\langle \xi^{a}\xi^{c}v_{j}^{b}(t') \rangle\rangle$$
$$+\frac{q_{s}q_{s'}}{c^{2}}\int d\textbf{r}' d\textbf{p}'
\int dt'
(\partial_{t}\partial_{\textbf{r}}^{c}G(t,t',\textbf{r},\textbf{r}')) \partial_{b,\textbf{p}}
\langle\langle \xi^{a}\xi'^{c}v_{j}^{b}(t') \rangle\rangle
+\frac{q_{s}q_{s'}}{c^{2}}
\varepsilon^{bcd}\varepsilon^{dfg}
\int d\textbf{r}' d\textbf{p}'
\int dt'
(\partial_{\textbf{r}}^{f}G(t,t',\textbf{r},\textbf{r}'))\times$$
$$\times\partial_{b,\textbf{p}}
\langle\langle \xi^{a}v_{i}^{c}(t)v_{j}^{g}(t') \rangle\rangle
+\frac{q_{s}q_{s'}}{c^{2}}
\varepsilon^{bcd}\varepsilon^{dfg}
\int d\textbf{r}' d\textbf{p}'
\int dt'
(\partial_{\textbf{r}}^{f}\partial_{\textbf{r}}^{l}G(t,t',\textbf{r},\textbf{r}'))
\partial_{b,\textbf{p}}
\langle\langle \xi^{a}\xi^{l}v_{i}^{c}(t)v_{j}^{g}(t') \rangle\rangle$$
\begin{equation}\label{RHD2021ClLM}
-\frac{q_{s}q_{s'}}{c^{2}}
\varepsilon^{bcd}\varepsilon^{dfg}
\int d\textbf{r}' d\textbf{p}'
\int dt'
(\partial_{\textbf{r}}^{f}\partial_{\textbf{r}}^{l}G(t,t',\textbf{r},\textbf{r}'))
\partial_{b,\textbf{p}}
\langle\langle \xi^{a}\xi'^{l}v_{i}^{c}(t)v_{j}^{g}(t') \rangle\rangle
=0\end{equation}


The two-coordinate distribution functions are presented as the double brackets of the corresponding values with no usage of specific notations
since part of them are introduced above,
while part of them get no specific notation.
Here, we make transition to the self-consistent field approximation.
To this end, we split the two-coordinate distribution functions on the product of one-coordinate distribution functions
(we do it in the same way
as it is done above for the equation of evolution of the scalar distribution function)
$$\partial_{t}d^{a}
+\partial^{b}_{\textbf{r}} J_{D}^{ba}
+q_{s}(\partial_{b,\textbf{p}}d^{a})
\biggl[E^{b}_{ext}(\textbf{r},t)
-q_{s}\biggl(
\partial^{b}_{\textbf{r}}
\int d\textbf{r}' d\textbf{p}'
\int dt'
G(t,t',\textbf{r},\textbf{r}')
f(\textbf{r}',\textbf{p}',t')$$
$$+\frac{1}{c^{2}}
\int d\textbf{r}' d\textbf{p}'
\int dt'
\partial_{t}G(t,t',\textbf{r},\textbf{r}')
\langle v_{j}^{b}(t')\rangle(\textbf{r}',\textbf{p}')
\biggr)
+q_{s}\partial^{c}_{\textbf{r}}\biggl(
\partial^{b}_{\textbf{r}}
\int d\textbf{r}' d\textbf{p}'
\int dt'
G(t,t',\textbf{r},\textbf{r}')
d^{c}(\textbf{r}',\textbf{p}',t')$$
$$+\frac{1}{c^{2}}
\int d\textbf{r}' d\textbf{p}'
\int dt'
\partial_{t}G(t,t',\textbf{r},\textbf{r}')
J_{D}^{bc}(\textbf{r}',\textbf{p}',t')
\biggr)\biggr]
+q_{s}(\partial_{b,\textbf{p}}q^{ac})
\biggl[\partial_{c,\textbf{r}}E^{b}_{ext}(\textbf{r},t)$$
$$
-q_{s}\partial_{c,\textbf{r}}\biggl(
\partial^{b}_{\textbf{r}}
\int d\textbf{r}' d\textbf{p}'
\int dt'
G(t,t',\textbf{r},\textbf{r}')
f(\textbf{r}',\textbf{p}',t')
+\frac{1}{c^{2}}
\int d\textbf{r}' d\textbf{p}'
\int dt'
\partial_{t}G(t,t',\textbf{r},\textbf{r}')
\langle v_{j}^{b}(t')\rangle(\textbf{r}',\textbf{p}')
\biggr)\biggr]
+\frac{q_{s}}{c}\varepsilon^{bcd}(\partial_{b,\textbf{p}}J_{D}^{ca})\times$$
$$\times\biggl[B_{ext}^{d}+\frac{q_{s'}}{c}\varepsilon^{dfg}\partial_{f,\textbf{r}}
\biggl(
\int d\textbf{r}' d\textbf{p}'
\int dt'
G(t,t',\textbf{r},\textbf{r}')
\langle v_{j}^{g}(t')\rangle(\textbf{r}',\textbf{p}')
-\partial_{l,\textbf{r}}\int d\textbf{r}' d\textbf{p}'
\int dt'
G(t,t',\textbf{r},\textbf{r}')
J_{D}^{gl}(\textbf{r}',\textbf{p}',t')
\biggr)\biggr]$$
\begin{equation}\label{RHD2021ClLM dip moment Distrib Function evol SCF appr}
+\frac{q_{s}}{c}\varepsilon^{bcd}(\partial_{b,\textbf{p}}J_{Q}^{caf})
\biggl[\partial_{f,\textbf{r}}B_{ext}^{d}+\frac{q_{s'}}{c}\varepsilon^{dfg}\partial_{f,\textbf{r}}\partial_{l,\textbf{r}}
\int d\textbf{r}' d\textbf{p}'
\int dt'
G(t,t',\textbf{r},\textbf{r}')
\langle v_{j}^{g}(t')\rangle(\textbf{r}',\textbf{p}')\biggr]
=0.
\end{equation}

\end{widetext}

On this stage we can introduce the averaged scalar and vector potentials of the electromagnetic field
and corresponding representation of the integral terms within
$\textbf{E}_{int}$ and $\textbf{B}_{int}$.
This electromagnetic field obeys the Maxwell equations (\ref{RHD2021ClLM Maxwell div B and rot E}),
(\ref{RHD2021ClLM Maxwell rot B kin}), (\ref{RHD2021ClLM div E kin}).

Finally, equation (\ref{RHD2021ClLM dip moment Distrib Function evol SCF appr}) is represented in terms of $\textbf{E}_{int}$ and $\textbf{B}_{int}$:
$$\partial_{t}d^{a}+\partial_{\textbf{r}}^{b}J_{D}^{ba}+q_{s}(\partial_{\textbf{p}}^{b}d^{a})E^{b}
+q_{s}(\partial_{\textbf{p}}^{b}q^{ac})\partial_{\textbf{r}}^{c}E^{b}$$
\begin{equation}\label{RHD2021ClLM dip moment Distrib Function evol SCF appr FIN}
+\frac{q_{s}}{c}\varepsilon^{bcd}(\partial_{\textbf{p}}^{b}J^{ca}_{D})B^{d}
+\frac{q_{s}}{c}\varepsilon^{bcd}(\partial_{\textbf{p}}^{b}J^{caf}_{Q})\partial_{\textbf{r}}^{f}B^{d}=0. \end{equation}

\section{Equation for the distribution function of velocity}

Let us repeat the definition of the considering vector distribution function (\ref{RHD2022ClSCF function F def REL})
\begin{equation}\label{RHD2022ClSCF function F def REL REPEAT}
\textbf{j}(\textbf{r},\textbf{p},t)\equiv\langle\textbf{v}_{i}(t)\rangle\equiv\frac{1}{\Delta}\frac{1}{\Delta_{p}}
\int_{\Delta,\Delta_{p}}
d\mbox{\boldmath $\xi$}
d\mbox{\boldmath $\eta$}
\sum_{i=1}^{N/2} \textbf{v}_{i}(t)
\delta_{\textbf{r}i}
\delta_{\textbf{p}i},
\end{equation}

If we consider the evolution of function (\ref{RHD2022ClSCF function F def REL REPEAT})
we would have four terms in the initial form of the kinetic equation
$$\partial_{t}\textbf{j}(\textbf{r},\textbf{p},t)
=\partial_{t}\langle \textbf{v}_{i}(t)\rangle$$
\begin{equation}\label{RHD2021ClLM function F REL Evolution 1}
=\langle \dot{\textbf{v}}_{i}(t)\rangle -\partial_{r}^{b}\langle \textbf{v}_{i}(t)v_{i}^{b}(t)\rangle
- \partial_{p}^{b}\langle \textbf{v}_{i}(t)\dot{p}_{i}^{b}(t)\rangle. \end{equation}

However, we can represent the velocity of each particle $\textbf{v}_{i}(t)$ via parameters $\textbf{p}$ and $\mbox{\boldmath $\eta$}$
which do not depend on time (see equation (\ref{RHD2022ClSCF function F repr 1 REL}))
$$\partial_{t}\textbf{j}(\textbf{r},\textbf{p},t)
=\partial_{t}\langle \textbf{v}+\Delta\textbf{v}\rangle$$
\begin{equation}\label{RHD2021ClLM function F REL Evolution 2}
=-\partial_{r}^{b}\langle \textbf{v}_{i}(t)v_{i}^{b}(t)\rangle
- \partial_{p}^{b}\langle \textbf{v}_{i}(t)\dot{p}_{i}^{b}(t)\rangle. \end{equation}

Equation (\ref{RHD2021ClLM function F REL Evolution 1}) can be simplified to equation (\ref{RHD2021ClLM function F REL Evolution 1}) using
the kinetic equation for the scalar distribution function $f(\textbf{r},\textbf{p},t)$.

We consider equation (\ref{RHD2021ClLM function F REL Evolution 2}) in more details
$$\partial_{t}j^{a}(\textbf{r},\textbf{p},t)
+\partial_{r}^{b}\langle v_{i}^{a}(t)v_{i}^{b}(t)\rangle$$
$$+q_{s}\partial_{p}^{b}\langle v_{i}^{a}E_{i}^{b}(\textbf{r}+\mbox{\boldmath $\xi$})
+ \varepsilon^{bcd}v_{i}^{a}v_{i}^{c}B_{i}^{d}(\textbf{r}+\mbox{\boldmath $\xi$})\rangle$$
$$-q_{s}q_{s'}\int d\textbf{r}'d\textbf{p}'\int dt'
\langle\langle v_{i}^{a}(t) \partial_{r}^{b}G(t,t',\textbf{r}+\mbox{\boldmath $\xi$},\textbf{r}'+\mbox{\boldmath $\xi$}') \partial_{p}^{b}\rangle\rangle$$
$$-\frac{q_{s}q_{s'}}{c^{2}}\int d\textbf{r}'d\textbf{p}'\int dt'
\langle\langle v_{i}^{a}(t) v_{j}^{b}(t') \partial_{t} G(t,t',\textbf{r}+\mbox{\boldmath $\xi$},\textbf{r}'+\mbox{\boldmath $\xi$}') \partial_{p}^{b}\rangle\rangle$$
$$+\frac{q_{s}q_{s'}}{c^{2}}\varepsilon^{bcd}\varepsilon^{dfg} \int d\textbf{r}'d\textbf{p}'\int dt'\times$$
\begin{equation}\label{RHD2021ClLM function F REL Evolution 3 Large}
\times \langle\langle v_{i}^{a}(t)v_{i}^{c}(t)v_{j}^{g}(t')
\partial_{r}^{f} G(t,t',\textbf{r}+\mbox{\boldmath $\xi$},\textbf{r}'+\mbox{\boldmath $\xi$}') \partial_{p}^{b}\rangle\rangle =0.\end{equation}

Next, we represent equation (\ref{RHD2021ClLM function F REL Evolution 3 Large}) via the multipole expansion
$$\partial_{t}j^{a}(\textbf{r},\textbf{p},t)
+\partial_{r}^{b}\langle v_{i}^{a}(t)v_{i}^{b}(t)\rangle
+q_{s}\biggl(E^{b}(\textbf{r},t) \partial_{p}^{b}\langle v_{i}^{a} \rangle$$
$$+(\partial_{r}^{c}E^{b}) \partial_{p}^{b}\langle v_{i}^{a} \xi^{c} \rangle
+\varepsilon^{bcd}B^{d} \partial_{p}^{b}\langle v_{i}^{a}v_{i}^{c} \rangle
+\varepsilon^{bcd}(\partial^{f}B^{d}) \partial_{p}^{b}\langle v_{i}^{a}v_{i}^{c}\xi^{f} \rangle\biggr)$$
$$-q_{s}q_{s'}\int d\textbf{r}'d\textbf{p}'\int dt' \partial_{r}^{b}G(t,t',\textbf{r},\textbf{r}')
\partial_{p}^{b}\langle\langle v_{i}^{a}(t)\rangle\rangle$$
$$-q_{s}q_{s'}\int d\textbf{r}'d\textbf{p}'\int dt'  \partial_{r}^{b}\partial_{r}^{c}G(t,t',\textbf{r},\textbf{r}')
\partial_{p}^{b}\langle\langle \xi^{c} v_{i}^{a}(t)\rangle\rangle$$
$$+q_{s}q_{s'}\int d\textbf{r}'d\textbf{p}'\int dt'  \partial_{r}^{b}\partial_{r}^{c}G(t,t',\textbf{r},\textbf{r}')
\partial_{p}^{b}\langle\langle \xi'^{c}v_{i}^{a}(t)\rangle\rangle$$
$$-\frac{q_{s}q_{s'}}{c^{2}}\int d\textbf{r}'d\textbf{p}'\int dt'  \partial_{t}G(t,t',\textbf{r},\textbf{r}')
\partial_{p}^{b}\langle\langle v_{i}^{a}(t)v_{j}^{b}(t')\rangle\rangle$$
$$+\frac{q_{s}q_{s'}}{c^{2}} \varepsilon^{bcd}\varepsilon^{dfg}\int d\textbf{r}'d\textbf{p}'\int dt'   \times$$
$$\times\partial_{r}^{f}G(t,t',\textbf{r},\textbf{r}')
\partial_{p}^{b}\langle\langle v_{i}^{a}(t)v_{i}^{c}(t)v_{j}^{g}(t')\rangle\rangle $$
$$+\frac{q_{s}q_{s'}}{c^{2}} \varepsilon^{bcd}\varepsilon^{dfg}\int d\textbf{r}'d\textbf{p}'\int dt'   \times$$
\begin{equation}\label{RHD2021ClLM function F REL Evolution 4 MultiPole appr}
\times\partial_{r}^{f}\partial_{r}^{h}G(t,t',\textbf{r},\textbf{r}')
\partial_{p}^{b}\langle\langle v_{i}^{a}(t)v_{i}^{c}(t)v_{j}^{g}(t')\xi^{h}\rangle\rangle =0.
\end{equation}

Finally, we present equation (\ref{RHD2021ClLM function F REL Evolution 4 MultiPole appr}) in the self-consistent field approximation
$$\partial_{t}j^{a}(\textbf{r},\textbf{p},t)
+\partial_{r}^{b}\langle v_{i}^{a}(t)v_{i}^{b}(t)\rangle
+q_{s}\biggl(E^{b}_{ext}(\textbf{r},t) \partial_{p}^{b}\langle v_{i}^{a} \rangle$$
$$+(\partial_{r}^{c}E^{b}_{ext}) \partial_{p}^{b}\langle v_{i}^{a} \xi^{c} \rangle
+\varepsilon^{bcd}(B^{d}_{ext} \partial_{p}^{b}\langle v_{i}^{a}v_{i}^{c} \rangle
+(\partial^{f}B^{d}) \partial_{p}^{b}\langle v_{i}^{a}v_{i}^{c}\xi^{f} \rangle)\biggr)$$
$$-q_{s}q_{s'}(\partial_{p}^{b}j^{a})
\partial_{r}^{b}\int d\textbf{r}'d\textbf{p}'\int dt'G(t,t',\textbf{r},\textbf{r}')
f(\textbf{r}',\textbf{p}',t')$$
$$-q_{s}q_{s'}(\partial_{p}^{b}J_{D}^{ac})
\partial_{r}^{b}\partial_{r}^{c}\int d\textbf{r}'d\textbf{p}'\int dt'G(t,t',\textbf{r},\textbf{r}')
f(\textbf{r}',\textbf{p}',t')$$
$$+q_{s}q_{s'} (\partial_{p}^{b}j^{a})
\partial_{r}^{b}\partial_{r}^{c}\int d\textbf{r}'d\textbf{p}'\int dt'G(t,t',\textbf{r},\textbf{r}')
d^{c}(\textbf{r}',\textbf{p}',t')$$
$$-\frac{q_{s}q_{s'}}{c^{2}}(\partial_{p}^{b}j^{a})
\partial_{t}\int d\textbf{r}'d\textbf{p}'\int dt'G(t,t',\textbf{r},\textbf{r}')
j^{c}(\textbf{r}',\textbf{p}',t')$$
$$+\frac{q_{s}q_{s'}}{c^{2}} \varepsilon^{bcd}\varepsilon^{dfg}
(\partial_{p}^{b}\langle v_{i}^{a}(t)v_{i}^{c}(t)\rangle)\times$$
$$\times\partial_{r}^{f}\int d\textbf{r}'d\textbf{p}'\int dt'G(t,t',\textbf{r},\textbf{r}')
j^{g}(\textbf{r}',\textbf{p}',t') $$
$$+\frac{q_{s}q_{s'}}{c^{2}} \varepsilon^{bcd}\varepsilon^{dfg}
(\partial_{p}^{b}\langle v_{i}^{a}(t)v_{i}^{c}(t)\xi^{h}\rangle)\times$$
\begin{equation}\label{RHD2021ClLM function F REL Evolution 5 SCF appr}
\times\partial_{r}^{f}\partial_{r}^{h}\int d\textbf{r}'d\textbf{p}'\int dt'G(t,t',\textbf{r},\textbf{r}')
j^{g}(\textbf{r}',\textbf{p}',t') =0,
\end{equation}
where
\begin{equation}\label{RHD2021ClLM}
\langle\langle v_{i}^{a}(t)\rangle\rangle(\textbf{r},\textbf{r}',\textbf{p},\textbf{p}',t,t')
=j^{a}(\textbf{r},\textbf{p},t)\cdot f(\textbf{r}',\textbf{p}',t') ,\end{equation}
\begin{equation}\label{RHD2021ClLM} \langle\langle \xi^{c} v_{i}^{a}(t)\rangle\rangle
=J_{D}^{ac}(\textbf{r},\textbf{p},t)\cdot f(\textbf{r}',\textbf{p}',t'),\end{equation}
\begin{equation}\label{RHD2021ClLM} \langle\langle \xi'^{c}v_{i}^{a}(t)\rangle\rangle
=j^{a}(\textbf{r},\textbf{p},t)\cdot d^{c}(\textbf{r}',\textbf{p}',t') ,\end{equation}
\begin{equation}\label{RHD2021ClLM} \langle\langle v_{i}^{a}(t)v_{j}^{b}(t')\rangle\rangle
=j^{a}(\textbf{r},\textbf{p},t)\cdot j^{b}(\textbf{r}',\textbf{p}',t'),\end{equation}
\begin{equation}\label{RHD2021ClLM} \langle\langle v_{i}^{a}(t)v_{i}^{c}(t)v_{j}^{g}(t')\rangle\rangle
=\langle v_{i}^{a}(t)v_{i}^{c}(t)\rangle\cdot j^{g}(\textbf{r}',\textbf{p}',t'),\end{equation}
and
\begin{equation}\label{RHD2021ClLM}
\langle\langle v_{i}^{a}(t)v_{i}^{c}(t)v_{j}^{g}(t')\xi^{h}\rangle\rangle=
\langle v_{i}^{a}(t)v_{i}^{c}(t)\xi^{h}\rangle \cdot j^{g}(\textbf{r}',\textbf{p}',t').\end{equation}

Finally, we introduce electric $\textbf{E}_{int}$ and magnetic $\textbf{B}_{int}$ fields caused by the charges of the system
and present the kinetic equation for the velocity distribution function in compact form:
$$\partial_{t}j^{a}+\partial_{\textbf{r}}^{b}\langle v_{i}^{a}(t)v_{i}^{b}(t)\rangle
+q_{s}E^{b}\partial_{\textbf{r}}^{b}j^{a}
+q_{s}(\partial_{\textbf{r}}^{c}E^{b})\partial_{\textbf{p}}^{b}J_{D}^{ac}$$
\begin{equation}\label{RHD2021ClLM function F REL Evolution 5 SCF appr FIN}
+\frac{q_{s}}{c}\varepsilon^{bcd}[B^{d}\partial_{\textbf{p}}^{b}\langle v_{i}^{a}(t)v_{i}^{c}(t)\rangle
+(\partial^{f}B^{d}) \partial_{\textbf{p}}^{b}\langle v_{i}^{a}(t)v_{i}^{c}(t)\xi^{f}\rangle]=0. \end{equation}

\section{On the extended closed set of kinetic equations containing scalar and vector distribution functions}

Equations for evolution of $f(\textbf{r},\textbf{p},t)$, $d^{a}(\textbf{r},\textbf{p},t)$ and $j^{a}(\textbf{r},\textbf{p},t)$
(see equations (\ref{RHD2022ClSCF Vlasov eq multipole SCF appr}),
(\ref{RHD2021ClLM dip moment Distrib Function evol SCF appr FIN}),
(\ref{RHD2021ClLM function F REL Evolution 5 SCF appr FIN}))
give incomplete set of equations.
This set requires additional assumptions to make final truncation.

Let us repeat equations (\ref{RHD2022ClSCF Vlasov eq multipole SCF appr}),
(\ref{RHD2021ClLM dip moment Distrib Function evol SCF appr FIN}),
(\ref{RHD2021ClLM function F REL Evolution 5 SCF appr FIN})
together to make additional analysis  of these equations
$$\partial_{t}f+
\textbf{v}\cdot\nabla f
+\nabla\cdot \textbf{F}
+q_{s}\biggl(\textbf{E}+\frac{1}{c}\textbf{v}\times\textbf{B}\biggr)\cdot\frac{\partial f}{\partial \textbf{p}}$$
$$+q_{s}\biggl(\partial^{b}\textbf{E}+\frac{1}{c}\textbf{v}\times\partial^{b}\textbf{B}\biggr)
\cdot\frac{\partial d_{s}^{b}}{\partial \textbf{p}}
+q_{s}\biggl(\partial^{b}\partial^{c}\textbf{E}$$
$$
+\frac{q_{s}}{c}\textbf{v}\times\partial^{b}\partial^{c}\textbf{B}\biggr)\cdot\frac{\partial Q^{bc}}{\partial \textbf{p}}
+\frac{q_{s}}{c}\varepsilon^{abc}\biggl(B^{c}\cdot\partial_{a,\textbf{p}}F^{b}  $$
\begin{equation}\label{RHD2022ClSCF Vlasov eq multipole SCF appr b}
+\partial^{d}B^{c}\cdot \partial_{a,\textbf{p}}  j_{D}^{bd}
+\partial^{d}\partial^{f}B^{c}\cdot \partial_{a,\textbf{p}}   j_{Q}^{bdf}  \biggr)
=0, \end{equation}
and
$$\partial_{t}d^{a}+\partial_{\textbf{r}}^{b}J_{D}^{ba}+q_{s}(\partial_{\textbf{p}}^{b}d^{a})E^{b}
+q_{s}(\partial_{\textbf{p}}^{b}Q^{ac})\partial_{\textbf{r}}^{c}E^{b}$$
\begin{equation}\label{RHD2021ClLM dip moment Distrib Function evol SCF appr FIN b}
+\frac{q_{s}}{c}\varepsilon^{bcd}(\partial_{\textbf{p}}^{b}J^{ca}_{D})B^{d}
+\frac{q_{s}}{c}\varepsilon^{bcd}(\partial_{\textbf{p}}^{b}J^{caf}_{Q})\partial_{\textbf{r}}^{f}B^{d}=0, \end{equation}
and
$$\partial_{t}j^{a}+\partial_{\textbf{r}}^{b}\langle v_{i}^{a}(t)v_{i}^{b}(t)\rangle
+q_{s}E^{b}\partial_{\textbf{r}}^{b}j^{a}
+q_{s}(\partial_{\textbf{r}}^{c}E^{b})\partial_{\textbf{p}}^{b}J_{D}^{ac}$$
\begin{equation}\label{RHD2021ClLM function F REL Evolution 5 SCF appr FIN b}
+\frac{q_{s}}{c}\varepsilon^{bcd}[B^{d}\partial_{\textbf{p}}^{b}\langle v_{i}^{a}(t)v_{i}^{c}(t)\rangle
+(\partial^{f}B^{d}) \partial_{\textbf{p}}^{b}\langle v_{i}^{a}(t)v_{i}^{c}(t)\xi^{f}\rangle]=0. \end{equation}

Here we have equations for three distribution functions $f$, $d^{a}$, and $j^{a}=v^{a}f+F^{a}$,
but we have a number of other distribution functions
like $Q^{bc}$, $J_{D}^{bc}$, $J_{Q}^{bcd}$,
$\langle v_{i}^{a}(t)v_{i}^{c}(t)\rangle$, and $\langle v_{i}^{a}(t)v_{i}^{c}(t)\xi^{f}\rangle$.
We need to obtain the closed set of equations at this step.
Before we make this truncation, we need to provide discussion of physical picture hidden in these functions.


\begin{figure}\includegraphics[width=8cm,angle=0]{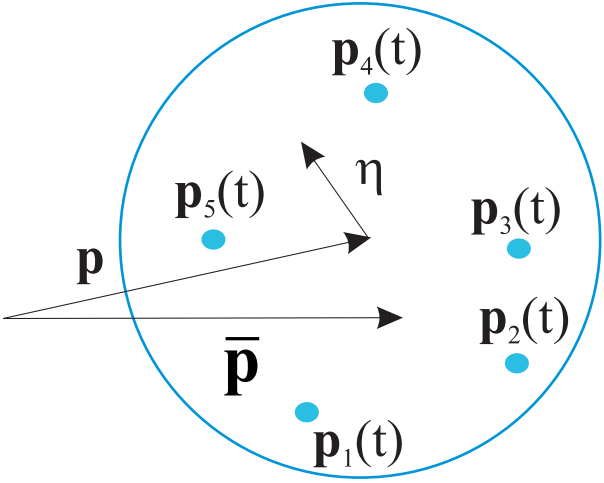}
\caption{\label{RHD2022ClSCF Fig 02}
The illustration of the $\Delta_{\textbf{p}}$-vicinity in the momentum space with few numbered particles being in the vicinity.
Here, $\textbf{p}$ is the center of the vicinity,
$\mbox{\boldmath $\eta$}$ is the vector scanning the vicinity,
$\textbf{p}_{i}(t)$ are the momentums of particles,
$\bar{\textbf{p}}$ is the average value of the momentum of particles being in the vicinity,
while vector $\bar{\textbf{p}}$ differs from the center of the vicinity $\textbf{p}$.}
\end{figure}

\subsection{Kinetic theory of fluctuations}

Considering transition from the kinetic theory
(the field form of the classical mechanics in the six dimensional space)
to the hydrodynamic theory
(the field form of the classical mechanics in the three dimensional space)
we get the concentration of particles
$n(\textbf{r},t)=\int f(\textbf{r},\textbf{p},t)d\textbf{p}$,
the current of particles
$\textbf{j}(\textbf{r},t)=n(\textbf{r},t)\textbf{v}(\textbf{r},t)$
$=\int \textbf{v}f(\textbf{r},\textbf{p},t)d\textbf{p}$.
Difference $\textbf{v}-\textbf{v}(\textbf{r},t)$ is the measure of chaotic motion,
which can be associated with the thermal motion.
This interpretation getting closer to the thermodynamic temperature
if the system get though the process of thermalization.
For the quantum systems we also need to include other than thermal mechanisms
for the symmetric distribution of particles over the quantum states in the momentum space.
Major contribution is given by the Pauli blocking existing in systems of fermions.

The average velocity $
\int \textbf{v}f(\textbf{r},\textbf{p},t)d\textbf{p}$ gives the velocity field.
The deviation from the average velocity
$\textbf{v}-\textbf{v}(\textbf{r},t)$ leads to other hydrodynamic functions,
like the temperature scalar field, the pressure tensor field, etc.
These functions are also smooth functions
which are defined on certain scale.
It leaves as the question:
how fluctuations can appear in our model?
Since the fluctuations is the natural phenomenon
following from the individual motion of particles governed be the microscopic equations.

Basically, the chaotic motion of particles gives small variation of macroscopic functions
due to the continuous unequal exchange of particles between nearest vicinities.
Hence, the concentration
$n(\textbf{r},t)=\int f(\textbf{r},\textbf{p},t)d\textbf{p}$
(\ref{RHD2022ClSCF concentration definition}),
the distribution function
$f(\textbf{r},\textbf{p},t)$
(\ref{RHD2022ClSCF distribution function definition}), etc
show small variation over time and coordinate.
These fluctuations can be considered via the correlations
like space correlations $\overline{n(\textbf{r},t)n(\textbf{r}',t)}$
or the time correlations $\widetilde{n(\textbf{r},t)n(\textbf{r},t')}$,
where we introduce additional average on the macroscopic space scale
$\bar{A}=(1/\Delta V)\int_{\Delta V}A dV$,
or the average over the interval of time $\Delta T$:
$\tilde{A}=(1/\Delta T)\int_{\Delta T}A dt$.

The distribution function is constructed on a certain scale in the coordinate and momentum space.
If we want to trace the deviation from average
(particularly in the momentum space)
we need to focus our attention on the deviation of momentum of all particles in $\Delta$-vicinity from the middle point of the vicinity
$\mbox{\boldmath $\eta$}$.
Parameter $\Delta\textbf{v}_{i}(t)$ is also associated with $\mbox{\boldmath $\eta$}$.
If we imagine a $\Delta$-vicinity with 5 particles
(it is an imaginary example, for the illustration, see Fig. \ref{RHD2022ClSCF Fig 02},
real vicinity contain a large number of particles)
we can count the average momentum
$\bar{\textbf{p}}=(1/5)\sum_{i=1}^{5}\textbf{p}_{i}(t)\neq \textbf{p}$
and we get a deviation from $\textbf{p}$
(the moment of the center of the vicinity).
However, the distribution function $f(\textbf{r},\textbf{p},t)$ count all these particles as the particles with momentum $\textbf{p}$.
In other words $\sum_{i=1}^{5}\mbox{\boldmath $\eta$}_{i}\neq 0$
(in general $\mbox{\boldmath $\eta$}$ is a parameter scanning the vicinity,
but here we introduce particular $\mbox{\boldmath $\eta$}_{i}$ as
$\mbox{\boldmath $\eta$}_{i}(t)=\textbf{p}_{i}(t)-\textbf{p}$
the deviation of momentum of particle being in the vicinity from the momentum of the center of vicinity).
This example shows that physical nature of fluctuations can be presented via functions
$\langle\eta^{a}\rangle$ and $\langle\eta^{a}\eta^{b}\rangle$
(or $\langle \Delta v_{i}^{a}(t)\rangle$
and $\langle \Delta v_{i}^{a}(t)\Delta v_{i}^{b}(t)\rangle$).
It explains necessity of the account of these functions in our model.
We consider equation for evolution of function $\langle \Delta v_{i}^{a}(t)\rangle$,
while function $\langle \Delta v_{i}^{a}(t)\Delta v_{i}^{b}(t)\rangle$ is assumed to be approximately expressed
via functions $f(\textbf{r},\textbf{p},t)$ and $\langle \Delta v_{i}^{a}(t)\rangle$.
Here, we introduce equation of state in order to make truncation of the set of kinetic equations
(in addition to the self-consistent field approximation).

In the chosen moment of time $t$ we have $\langle \Delta \textbf{v}_{i}\rangle \neq0$.
However, the fluctuations happen on some time scale $\tau$,
which can be associated with the time of propagation of average particle via the $\Delta$-vicinity:
$\tau=\sqrt[3]{\Delta}/v_{0}$,
where the average velocity of the $\Delta_{p}$-vicinity is
$v_{0}=pc/\sqrt{p^{2}+m_{s}^{2}c^{2}}$.
We can go further and make averaging of the kinetic equation over this time interval
$\tilde{a}=(1/\tau)\int_{0}^{\tau}adt$.
Here, we find
$\widetilde{\langle \Delta \textbf{v}_{i}\rangle} =0$,
but
$\widetilde{\langle \Delta v_{i}^{a} \Delta v_{i}^{b}\rangle} \neq0$.
As an estimation
we can choose
$\Delta \textbf{v}_{i,max}=\sqrt[3]{\Delta_{v}}$
$=\sqrt[3]{\Delta_{p}}c/\sqrt{\sqrt[3]{\Delta_{v}}^{2}+m_{s}^{2}c^{2}}$.



\subsection{Truncation method}

Let us consider simplified form of the kinetic equation for the vector distribution function of velocity
(\ref{RHD2021ClLM function F REL Evolution 5 SCF appr}).
Function $j^{a}(\textbf{r},\textbf{p},t)$ can be represented via two other functions introduced above
$j^{a}(\textbf{r},\textbf{p},t)=v^{a}\cdot f(\textbf{r},\textbf{p},t)$
$+F^{a}(\textbf{r},\textbf{p},t)$.
Equation (\ref{RHD2021ClLM function F REL Evolution 5 SCF appr}) includes function
$\langle v_{i}^{a}(t)v_{i}^{b}(t)\rangle$
$=\langle (v^{a}+\Delta v_{i}^{a})(v^{b}+\Delta v_{i}^{b})\rangle$
$=v^{a}v^{b}\langle 1\rangle +v^{a}\langle\Delta v_{i}^{b}\rangle +v^{b}\langle\Delta v_{i}^{a}\rangle$
$+\langle\Delta v_{i}^{a}\Delta v_{i}^{b}\rangle$
$=v^{a}v^{b}f(\textbf{r},\textbf{p},t)$
$+v^{a}F^{b}(\textbf{r},\textbf{p},t)$
$+v^{b}F^{a}(\textbf{r},\textbf{p},t)$
$+\langle\Delta v_{i}^{a}\Delta v_{i}^{b}\rangle$.
Here we extract the velocity from functions
$j_{s}^{a}$, $j_{D,s}^{ab}$, $\langle v_{i}^{a}(t)v_{i}^{b}(t)\rangle$, and $j_{Q,s}^{abc}$
to show the Lorentz force in more familiar way.
It helps us to make the truncation as well.
Finally, we represent
$\langle v_{i}^{a}(t)v_{i}^{b}(t)\xi^{c}\rangle$
$=v^{a}v^{b}\langle \xi^{c}\rangle +v^{a}\langle\Delta v_{i}^{b} \xi^{c}\rangle +v^{b}\langle\Delta v_{i}^{a}\xi^{c}\rangle$
$+\langle\Delta v_{i}^{a}\Delta v_{i}^{b} \xi^{c}\rangle$.

Our model includes the multipole moments in the coordinate and momentum space, like $d^{a}$ and $F^{a}$.
The multipole moments are defined on the scale of the $\Delta$-vicinity.
Therefore, they appear to be small.
On the macroscopic scale this value tends to zero.
Hence, we can introduce the small parameter $\varepsilon$ to indicate the relative value of these functions.
We find $d^{a}\sim\varepsilon f$, $F^{a}\sim\varepsilon f$, $Q^{ab}\sim\varepsilon^{2} f$,
$j^{ab}\sim\varepsilon^{2} f$, etc.
To make the truncation we need to estimate the high-rank tensor distribution functions included in the model
$Q^{ab}=\langle\xi^{a}\xi^{b}\rangle=\delta^{ab}\sqrt[3]{\Delta^{2}}f(\textbf{r},\textbf{p},t)$
$\langle \Delta v^{a}\Delta v^{b}\rangle=\delta^{ab}\sqrt[3]{\Delta_{v}^{2}}f(\textbf{r},\textbf{p},t)$,
and
$j_{D}^{ab}=\delta^{ab}\sqrt[3]{\Delta\Delta_{v}}f(\textbf{r},\textbf{p},t)$.
The third rank tensors are neglected.
Function $\langle \Delta v^{a}\Delta v^{b}\rangle$ is symmetric,
so it is reasonable to be proportional to $\delta^{ab}$.
However, function $j_{D}^{ab}$ is not symmetric,
but we assume that
deviations of same projections of the coordinate and momentum are correlated.
This correlation is included in the model via nonzero value of $j_{D}^{ab}$,
which can be traced in calculations of any specific problem.
Further comparison with experiment can show validity of our assumption.
If no correlation is found this function can be considered equal to zero in the future modification of the model.

We start the discussion of equations
(\ref{RHD2022ClSCF Vlasov eq multipole SCF appr b}),
(\ref{RHD2021ClLM dip moment Distrib Function evol SCF appr FIN b}),
(\ref{RHD2021ClLM function F REL Evolution 5 SCF appr FIN b})
with the first of them
(\ref{RHD2022ClSCF Vlasov eq multipole SCF appr b}).
We can extract terms proportional to the zero degree of the small parameter $\varepsilon$:
\begin{equation}\label{RHD2021ClLM eq partial f 1} \partial_{t}f+
\textbf{v}\cdot\nabla f
+q_{s}\biggl(\textbf{E}+\frac{1}{c}\textbf{v}\times\textbf{B}\biggr)\cdot\frac{\partial f}{\partial \textbf{p}}, \end{equation}
which correspond to the well-known Vlasov equation.
Next, we present terms proportional to the first degree of the small parameter $\varepsilon$:
\begin{equation}\label{RHD2021ClLM eq partial f 2}
q_{s}\biggl(\partial^{b}\textbf{E}+\frac{1}{c}\textbf{v}\times\partial^{b}\textbf{B}\biggr)
\cdot\frac{\partial d_{s}^{b}}{\partial \textbf{p}}
+\nabla \cdot\textbf{F}
+\frac{q_{s}}{c}\varepsilon^{abc}B^{c}\cdot\partial_{a,\textbf{p}}F^{b}.
 \end{equation}
Other terms in equation (\ref{RHD2022ClSCF Vlasov eq multipole SCF appr b})
correspond to the second and third degree of the small parameter $\varepsilon$.
They are neglected since we include the first correction to the Vlasov equation.

Next we consider equation (\ref{RHD2021ClLM dip moment Distrib Function evol SCF appr FIN b}).
Lowest order on parameter $\varepsilon$ is equal to one in this equation.
We consider the lowest order and one correction to it.
Let us show term existing in the lowest order:
\begin{equation}\label{RHD2021ClLM eq partial d 1}
\partial_{t}d^{a}+(\textbf{v}\cdot\nabla)d^{a}
+q_{s}(\partial_{\textbf{p}}^{b}d^{a})E^{b}
+\frac{q_{s}}{c}\varepsilon^{bcd}v^{c}(\partial_{\textbf{p}}^{b}d^{a})B^{d}, \end{equation}
and in the next order
$$\partial_{\textbf{r}}^{b}j_{D}^{ba}
+\frac{q_{s}}{c}\varepsilon^{bcd}(\partial_{\textbf{p}}^{b}j^{ca}_{D})B^{d}$$
\begin{equation}\label{RHD2021ClLM eq partial d 2}
+q_{s}(\partial_{\textbf{p}}^{b}Q^{ac})\partial_{\textbf{r}}^{c}E^{b}
+\frac{q_{s}}{c}\varepsilon^{bcd}v^{c}(\partial_{\textbf{p}}^{b}Q^{af})\partial_{\textbf{r}}^{f}B^{d}
. \end{equation}

Let us consider
equation (\ref{RHD2021ClLM function F REL Evolution 5 SCF appr FIN b}).
Function $j^{a}$ splits on two terms $v^{a}f+F^{a}$.
Obviously, equation (\ref{RHD2021ClLM function F REL Evolution 5 SCF appr FIN b}) contains contribution of terms
(\ref{RHD2021ClLM eq partial f 1}) and (\ref{RHD2021ClLM eq partial f 2}) multiplied by $v^{a}$.
Hence, all these terms are equation to zero in accordance with the equation of evolution of the scalar distribution function.
After this simplification we find that
the lowest order of terms in equation (\ref{RHD2021ClLM function F REL Evolution 5 SCF appr FIN b}) on the parameter $\varepsilon$
is equal to one.
It contains the following terms:
\begin{equation}\label{RHD2021ClLM}
\partial_{t}F^{a}+(\textbf{v}\cdot\nabla)F^{a}+q_{s}E^{b}\partial_{\textbf{p}}^{b}F^{a}
+\frac{q_{s}}{c}\varepsilon^{bcd}v^{c}B^{d}\partial_{\textbf{p}}^{b}F^{a}\end{equation}
In the next order on $\varepsilon$ we obtain
$$\partial_{\textbf{r}}^{b}\langle\Delta v^{a}\Delta v^{b}\rangle
+q_{s}(\partial_{\textbf{r}}^{c}E^{b})\partial_{p}^{b}j_{D}^{ac}$$
\begin{equation}\label{RHD2021ClLM}
+\frac{q_{s}}{c}\varepsilon^{bcd}B^{d}\partial_{\textbf{p}}^{b}\langle\Delta v^{a}\Delta v^{c}\rangle
+\frac{q_{s}}{c}\varepsilon^{bcd}v^{c}(\partial_{\textbf{r}}^{f}B^{d})\partial_{\textbf{p}}^{b}j_{D}^{af}.\end{equation}

\subsection{Final set of kinetic equations}

Let us combine together final parts of equations
(\ref{RHD2022ClSCF Vlasov eq multipole SCF appr b}),
(\ref{RHD2021ClLM dip moment Distrib Function evol SCF appr FIN b}),
(\ref{RHD2021ClLM function F REL Evolution 5 SCF appr FIN b})
with the equations of state
$$\partial_{t}f+
\textbf{v}\cdot\nabla f
+q_{s}\biggl(\textbf{E}+\frac{1}{c}\textbf{v}\times\textbf{B}\biggr)\cdot\frac{\partial f}{\partial \textbf{p}}$$
\begin{equation}\label{RHD2022ClSCF Vlasov eq multipole SCF appr c}
+q_{s}\biggl(\partial^{b}\textbf{E}+\frac{1}{c}\textbf{v}\times\partial^{b}\textbf{B}\biggr)
\cdot\frac{\partial d_{s}^{b}}{\partial \textbf{p}}
+\nabla \cdot\textbf{F}
+\frac{q_{s}}{c}\varepsilon^{abc}B^{c}\cdot\partial_{a,\textbf{p}}F^{b}=0,
 \end{equation}
$$\partial_{t}d^{a}+(\textbf{v}\cdot\nabla)d^{a}
+q_{s}(\partial_{\textbf{p}}^{b}d^{a})E^{b}
+\frac{q_{s}}{c}\varepsilon^{bcd}v^{c}(\partial_{\textbf{p}}^{b}d^{a})B^{d} $$
$$+\sqrt[3]{\Delta\Delta_{v}}\partial_{\textbf{r}}^{a}f
-\frac{q_{s}}{c}\sqrt[3]{\Delta\Delta_{v}}\varepsilon^{abd}(\partial_{\textbf{p}}^{b}f)B^{d}$$
\begin{equation}\label{RHD2021ClLM dip moment Distrib Function evol SCF appr FIN c}
+q_{s}\sqrt[3]{\Delta^{2}}(\partial_{\textbf{p}}^{b}f)\partial_{\textbf{r}}^{a}E^{b}
+\frac{q_{s}}{c}\sqrt[3]{\Delta^{2}}\varepsilon^{bcd}v^{c}(\partial_{\textbf{p}}^{b}f)\partial_{\textbf{r}}^{a}B^{d}=0,
\end{equation}
and
$$\partial_{t}F^{a}+(\textbf{v}\cdot\nabla)F^{a}+q_{s}E^{b}\partial_{\textbf{p}}^{b}F^{a}
+\frac{q_{s}}{c}\varepsilon^{bcd}v^{c}B^{d}\partial_{\textbf{p}}^{b}F^{a}$$
$$+\sqrt[3]{\Delta_{v}^{2}}\partial_{\textbf{r}}^{a}f
+q_{s}\sqrt[3]{\Delta\Delta_{v}}(\partial_{\textbf{r}}^{a}E^{b})\partial_{\textbf{p}}^{b}f$$
\begin{equation}\label{RHD2021ClLM function F REL Evolution 5 SCF appr FIN c}
-\frac{q_{s}}{c}\sqrt[3]{\Delta_{v}^{2}}\varepsilon^{abd}B^{d}\partial_{\textbf{p}}^{b}f
+\frac{q_{s}}{c}\sqrt[3]{\Delta\Delta_{v}}
\varepsilon^{bcd}v^{c}(\partial_{\textbf{r}}^{a}B^{d})\partial_{\textbf{p}}^{b}f=0.\end{equation}

The electromagnetic field in equations
(\ref{RHD2022ClSCF Vlasov eq multipole SCF appr c}),
(\ref{RHD2021ClLM dip moment Distrib Function evol SCF appr FIN c}),
(\ref{RHD2021ClLM function F REL Evolution 5 SCF appr FIN c}) obeys
the following Maxwell equations
\begin{equation}\label{RHD2021ClLM Maxwell div B and rot E truncated}
\begin{array}{cc}
  \nabla\cdot\textbf{B}_{int}=0, & \nabla\times \textbf{E}_{int}=-\frac{1}{c}\partial_{t}\textbf{B}_{int},
\end{array} \end{equation}
$$(\nabla\times \textbf{B}_{int})^{a}=\frac{1}{c}\partial_{t}E_{int} ^{a}+$$
$$+\frac{4\pi}{c}\sum_{s}^{} q_{s}\biggl(
\int
v^{a}f_{s}(\textbf{r},\textbf{p},t)
d\textbf{p}
+\partial^{b}\int
v^{a}d_{s}^{b}(\textbf{r},\textbf{p},t)
d\textbf{p}$$
\begin{equation}\label{RHD2021ClLM Maxwell rot B kin truncated}
+\int
F_{s}^{a}(\textbf{r},\textbf{p},t)
d\textbf{p}
+\sqrt[3]{\Delta\Delta_{v}}\partial^{a}\int
f(\textbf{r},\textbf{p},t)
d\textbf{p}
\biggr),
\end{equation}
$$\nabla\cdot \textbf{E}_{int}=4\pi \sum_{s}^{} q_{s}\biggl(\int f_{s}(\textbf{r},\textbf{p},t)d\textbf{p}$$
\begin{equation}\label{RHD2021ClLM div E kin truncated}
+\partial^{b}\int d_{s}^{b}(\textbf{r},\textbf{p},t)d\textbf{p}
+\frac{1}{2} \sqrt[3]{\Delta^{2}}\partial^{b}\partial^{b}\int f(\textbf{r},\textbf{p},t)
d\textbf{p}
+...\biggr). \end{equation}
which appear as corresponding modification of equations
(\ref{RHD2021ClLM div E kin}),
(\ref{RHD2021ClLM Maxwell div B and rot E}),
and
(\ref{RHD2021ClLM Maxwell rot B kin}).

\section{The spin-electron-acoustic waves propagation in the spin polarized electron gas of high density}


The degenerate macroscopically motionless
(being in the equilibrium state)
electron gas are described via the Vlasov kinetic equation for each spin projection of electrons \cite{Andreev PP 16 SSE kin}
\begin{equation}\label{RHD2021ClLM Vlasov for waves} \partial_{t}f_{s}+
\textbf{v}\cdot\nabla f_{s}
+q_{s}\biggl(\textbf{E}+\frac{1}{c}\textbf{v}\times\textbf{B}\biggr)\cdot\frac{\partial f_{s}}{\partial \textbf{p}}=0. \end{equation}
Subindex $s$ corresponds to the electrons with the spin-up $s=\uparrow$ or spin-down $s=\downarrow$.
We consider small perturbations of the equilibrium state $f_{s}=f_{0s}+\delta f_{s}$,
where $f_{0s}$ is the equilibrium distribution function,
and $\delta f_{s}$ is the perturbation of the distribution function,
which is chosen as the plane wave in the coordinate space $\delta f_{s}= F_{s}e^{-\imath\omega t+\imath k_{z}z}$,
with the constant amplitude $F_{s}$.
It lead to the linearized kinetic equation
\begin{equation}\label{RHD2021ClLM Vlasov for waves linear no B0} -\imath(\omega-k_{z}v_{z})\delta f_{s}
+q_{s}\delta \textbf{E}\cdot\frac{\partial f_{0s}}{\partial \textbf{p}}=0,\end{equation}
where $\delta \textbf{B}=0$ for the longitudinal waves.

Equilibrium distribution functions for each subspecies of electrons is chosen as the Fermi step
$f_{0s}=\theta(p_{Fs}-p)/(2\pi\hbar)^{3}$,
where $p_{Fs}=(6\pi n_{0s})^{1/3}\hbar$ is the radius of the Fermi sphere in the momentum space for species $s$.

We consider the longitudinal waves,
hence the perturbation of the electric field is parallel to the direction of the wave propagation $\textbf{k}\parallel \delta \textbf{E}$.

Equation (\ref{RHD2021ClLM Vlasov for waves linear no B0}) give expression for the perturbation of the distribution function of electrons
\begin{equation}\label{RHD2021ClLM perturb of f s} \delta f_{s}=-\imath q_{s}(\textbf{v} \cdot\delta \textbf{E})
\frac{\partial f_{0s}}{\partial\varepsilon} \frac{1}{\omega-k_{z}v_{z}}, \end{equation}
where
\begin{equation}\label{RHD2021ClLM} \frac{\partial f_{0s}}{\partial\varepsilon}
=\frac{\partial f_{0s}}{\partial p}\frac{\partial p}{\partial\varepsilon}
=-\frac{1}{(2\pi\hbar)^3}\delta(p-p_{Fs}) \frac{p c^{2}}{\varepsilon} .\end{equation}

Expression (\ref{RHD2021ClLM perturb of f s}) allows us to calculate the perturbations of concentration of the degenerate electrons
$$\delta n_{s}=\int\delta f_{s}(\textbf{r},\textbf{p},t) d\textbf{p}$$
\begin{equation}\label{RHD2021ClLM delta n via delta f}   =-\frac{2\pi\imath q_{s}p_{Fs}^{2}\delta E_{z}}{(2\pi\hbar)^3 k_{z}v_{Fs}}
\biggl[2+ \frac{\omega}{k_{z}v_{Fs}}ln\biggl(\frac{\omega-k_{z}v_{Fs}}{\omega+k_{z}v_{Fs}}\biggr)  \biggr]. \end{equation}

It leads to the dispersion equation
\begin{equation}\label{RHD2021ClLM dispersion equation LW and SEAW}
1+\frac{3}{2} \frac{\omega_{Le}^{2}}{k_{z}^{2}c^{2}}\sum_{s=u,d}\frac{n_{0s}}{n_{0e}}\gamma_{Fs}\frac{m_{s}^{2}c^{2}}{p_{Fs}^{2}}
\biggl[2+ \frac{\omega}{k_{z}v_{Fs}}ln\biggl(\frac{\omega-k_{z}v_{Fs}}{\omega+k_{z}v_{Fs}}\biggr)  \biggr]=0. \end{equation}

Presented dispersion equation is found for the motionless ions,
but we have two species of particles: spin-up electrons and spin-down electrons.
Both species of electrons are degenerate,
but they have different equilibrium concentrations.
So we have nonzero equilibrium spin polarization.
The spin polarization is caused by the external magnetic field.
However, it is not included in equation (\ref{RHD2021ClLM Vlasov for waves linear no B0}),
since is does not affect the final result (\ref{RHD2021ClLM perturb of f s}) for the waves propagating parallel to the magnetic field.

Equation (\ref{RHD2021ClLM dispersion equation LW and SEAW}) gives two solutions:
the Langmuir wave and the SEAW.
The SEAW has been studied in different regimes and using different method.
It is considered within nonrelativistic hydrodynamics \cite{Andreev PRE 15},
nonrelativistic kinetics \cite{Andreev PP 16 SSE kin},
and relativistic hydrodynamics \cite{Andreev 2112}.
Here, we present relativistic kinetic theory of the SEAW.
Numerical analysis of equation (\ref{RHD2021ClLM dispersion equation LW and SEAW}) shows
that the dispersion dependencies of the Langmuir wave and the SEAW getting close to each other at the large wave vectors,
which leads to an instability in the short-wavelength limit.
However, it can be subject of another work specified on the instabilities in degenerate plasmas related to the SEAWs,
while this paper is focused on derivation of the extended kinetic model.

\section{Conclusion}

Detailed derivation of kinetic equations for the relativistic plasmas has been presented.
It concludes in the set of three kinetic equations (for each species of particles):
the Vlasov kinetic equation for the scalar distribution function,
the equation of evolution of the vector distribution function of the electric dipole moment
(the fluctuation of local center of mass of the particles being in the physically infinitesimal volume),
and
the equation of evolution of the vector distribution function of the velocity
(the fluctuation of average momentum of the particles being in the physically infinitesimal volume).
This closed set of equations appears as the result of truncation of the quasi-infinite set of kinetic equations
(number of degrees of freedom is restricted due to the finite number of particles N),
which extends via the account of two-, three- and more coordinate distribution functions,
and higher multipole moments of one-coordinate (along with two-, three- and more coordinate) distribution functions.

The method of derivation shows
the representation of the microscopic motion of the individual particles in terms of the macroscopic distribution functions.
It gives representation of deterministic classical mechanics in terms of evolution collective functions.
No probabilistic approaches are used for the derivation or interpretation of presented equations and found functions.

Equations have been found for the relativistic plasmas
meaning both the large temperatures of the system (large velocities of the chaotic motion)
and large velocities of the ordered motion.
The motion of particles with the velocities close to the speed of light requires
the account of full retarding potentials for the electromagnetic field created by particles.
It reflects in the full Maxwell equations obtained in the self-consistent field approximation.
The Maxwell equations also include the distribution function of the electric dipole moment,
 distribution function of the velocity, etc as the sources of the electromagnetic field.

The model is found for the hot plasmas,
but it can be used for the degenerate electron gas with the Fermi velocity close to the speed of light.
Hence, it is applied to the spin-electron acoustic waves in the extremely dense plasmas.
The SEAWs are considered in the regime
of account of the scalar distribution function governed by the Vlasov equation.



\section{DATA AVAILABILITY}

Data sharing is not applicable to this article as no new data were
created or analyzed in this study, which is a purely theoretical one.

\appendix

\section{Potentials of the electromagnetic field in the multipole approximation}


The application of the full retarding potentials for the description of the field created by particles
(\ref{RHD2022ClSCF varphi via Green function rel})
and
(\ref{RHD2022ClSCF A via Green function rel})
reveals in the full set of Maxwell equations found in the self-consistent field approximation
(\ref{RHD2021ClLM div E kin}),
(\ref{RHD2021ClLM Maxwell div B and rot E}),
and
(\ref{RHD2021ClLM Maxwell rot B kin}).
However, firstly it appears (in the self-consistent field approximation) in the kinetic equation as the integral terms
simplifies after splitting of the two-coordinate distribution functions on the products of the corresponding one-coordinate distribution functions.
These integral terms are shown in the text,
but we want to specify the structure of the scalar and vector potentials of the electromagnetic field caused by different distribution functions
since the Maxwell equations
(\ref{RHD2021ClLM div E kin}),
(\ref{RHD2021ClLM Maxwell div B and rot E}),
and
(\ref{RHD2021ClLM Maxwell rot B kin})
show the combined contribution of all terms.

Therefore, the partial scalar potentials are:
\begin{equation}\label{RHD2021ClLM} \varphi_{0}=\int d\textbf{r}' d\textbf{p}'
\int dt'
G(t,t',\textbf{r},\textbf{r}')
f(\textbf{r}',\textbf{p}',t'), \end{equation}

\begin{equation}\label{RHD2021ClLM} \varphi_{D}=-\partial^{c}_{\textbf{r}}
\int d\textbf{r}' d\textbf{p}'
\int dt'
G(t,t',\textbf{r},\textbf{r}')
d^{c}(\textbf{r}',\textbf{p}',t'), \end{equation}
and
\begin{equation}\label{RHD2021ClLM}
\varphi_{Q}=
\frac{1}{2}
\partial^{c}_{\textbf{r}}\partial^{b}_{\textbf{r}}
\int d\textbf{r}' d\textbf{p}'
\int dt'
G(t,t',\textbf{r},\textbf{r}')
q^{bc}(\textbf{r}',\textbf{p}',t'). \end{equation}

The partial vector potentials are
\begin{equation}\label{RHD2021ClLM} A^{a}_{0}=\frac{1}{c}
\int d\textbf{r}' d\textbf{p}'
\int dt'
G(t,t',\textbf{r},\textbf{r}')
j^{a}(\textbf{r}',\textbf{p}',t'), \end{equation}
\begin{equation}\label{RHD2021ClLM} A_{D}^{a}=-\frac{1}{c}\partial^{b}_{\textbf{r}}
\int d\textbf{r}' d\textbf{p}'
\int dt'
G(t,t',\textbf{r},\textbf{r}')
J_{D}^{ab}(\textbf{r}',\textbf{p}',t'), \end{equation}
and
\begin{equation}\label{RHD2021ClLM} A_{Q}^{a}=
\frac{1}{2}\frac{1}{c}
\partial^{c}_{\textbf{r}}\partial^{b}_{\textbf{r}}
\int d\textbf{r}' d\textbf{p}'
\int dt'
G(t,t',\textbf{r},\textbf{r}')
J_{Q}^{abc}(\textbf{r}',\textbf{p}',t'). \end{equation}

These potentials correspond to the Maxwell equations presented in the text
(\ref{RHD2021ClLM Maxwell div B and rot E}), (\ref{RHD2021ClLM div E kin}), (\ref{RHD2021ClLM Maxwell rot B kin}).



\begin{thebibliography}{17}

\bibitem{Klimontovich book} Yu. L. Klimontovich, \emph{Statistical Physics} Harwood, New York (1986).


\bibitem{Israel AP 79}
W. Israel, J. M. Stewart, Transient relativistic thermodynamics and kinetic theory,
Annals Phys. \textbf{118}, 341 (1979).

\bibitem{Romatschke IJMPE 10}
P. Romatschke, "New developments in relativistic viscous hydrodynamics",
Int. J. Mod. Phys. E \textbf{19}, 1 (2010).

\bibitem{Kovtun JP A 12}
P. Kovtun, "Lectures on hydrodynamic fluctuations in relativistic theories"
J. Phys. A: Math. Theor. \textbf{45}, 473001 (2012).

\bibitem{Denicol PRD 12}
G. S. Denicol, H. Niemi, E. Molnar, D. H. Rischke,
"Derivation of transient relativistic fluid dynamics from the Boltzmann equation"
Phys. Rev. D \textbf{85}, 114047 (2012).

\bibitem{Heinz ARNPS 13}
U. Heinz, R. Snellings, "Collective Flow and Viscosity in Relativistic Heavy-Ion Collisions",
Annu. Rev. Nucl. Part. Sci. \textbf{63}, 123 (2013).


\bibitem{Florkowski RPP 18}
W. Florkowski, M. P. Heller, M. Spalinski, "New theories of relativistic hydrodynamics in the LHC era",
Rep. Prog. Phys. \textbf{81}, 046001 (2018).


\bibitem{Denicol PRD 19}
G. S. Denicol, E. Molnar, H. Niemi, D. H. Rischke,
"Resistive dissipative magnetohydrodynamics from the Boltzmann-Vlasov equation",
Phys. Rev. D \textbf{99}, 056017 (2019).

\bibitem{Hakim book Rel Stat Phys} Remi Hakim,
Introduction to Relativistic Statistical Mechanics Classical and Quantum, World Scientific Publishing Co. Pte. Ltd.,  2011.

\bibitem{Drofa TMP 96}  M. A. Drofa, L. S. Kuz'menkov,
"Continual approach to multiparticle systems with long-range interaction. Hierarchy of macroscopic fields and physical consequences",
Theoretical and Mathematical Physics \textbf{108}, 849 (1996).

\bibitem{Kuzmenkov CM 15} L. S. Kuzmenkov, Theoretical Physics: Classical Mechanics (Nauka, Moscow, 2015) [in Russian].

\bibitem{Weinberg Gr 72} S. Weinberg, Gravitation and Cosmology (John Wiley and Sons, Inc., NewYork, 1972).


\bibitem{deGrootFoundations1972}
S.R. de Groot, L.G. Suttorp,
"Foundations of Electrodynamics" (North-Holland 1972).

\bibitem{Kuzelev UFN 99} M. V. Kuzelev, A. A. Rukhadze,
"On the quantum description of the linear kinetics of a collisionless plasma",
Phys. Usp. \textbf{42}, 603–605 (1999).

\bibitem{Andreev PIERS 2012} L. S. Kuz'menkov and P. A. Andreev, "Microscopic Classic Hydrodynamic and Methods of Averaging",
presented in PIERS Proceedings, p. 158, August 19-23, Moscow, Russia 2012.

\bibitem{Andreev 2021 05} P. A. Andreev, "On the structure of relativistic hydrodynamics for hot plasmas",
Phys. Scr. \textbf{97}, 085602 (2022).

\bibitem{Andreev Ph 15} P. A. Andreev,
"Quantum kinetics of spinning neutral particles: General theory and Spin wave dispersion"
Physica A \textbf{432}, 108 (2015).


\bibitem{Zubkov JPhCS 19} V. V. Zubkov, "Distribution functions for continuous medium
without probability hypotheses", J. Phys.: Conf. Ser. \textbf{1352}, 012067 (2019).

\bibitem{Zubkov 2006} V.V. Zubkov, A.V. Zubkova,
"Irreversibility in classical kinetic theory: retardation of interaction and distribution functions",
arXiv:2006.11565.

\bibitem{Zubkov VNSU 22} V. V. Zubkov, D. A. Mayfat, Yashkin K.Yu. Tensor field method in the linear response theory // Vestnik NovSU.
Issue: Engineering Sciences. 2022. № 3(128). P.21–25.
DOI: https://doi.org/10.34680/2076-8052.2022.3(128).21-25



\bibitem{Maksimov QHM 99} L. S. Kuz'menkov, S. G. Maksimov, "Quantum hydrodynamics of particle systems with
Coulomb interaction and quantum Bohm potential," Theor. Math. Phys. \textbf{118}, 227 (1999).

\bibitem{MaksimovTMP 2001} L. S. Kuz'menkov, S. G. Maksimov, V. V. Fedoseev,
"Microscopic quantum hydrodynamics of systems of
fermions: part I",
Theor. Math. Phys. \textbf{126}, 110 (2001).

\bibitem{Andreev LP 19} P. A. Andreev,
Hydrodynamic model of a Bose–Einstein condensate with anisotropic short-range interaction and bright solitons in a repulsive Bose–Einstein condensate,
Laser Phys. \textbf{29}, 035502 (2019).

\bibitem{Andreev LP 21 fermions} P. A. Andreev,
"Extended hydrodynamics of degenerate partially spin polarized fermions with
short-range interaction up to the third order by interaction radius approximation",
Laser Phys. \textbf{31}, 045501 (2021).

\bibitem{Andreev LPL 21} P. A. Andreev,
"Hydrodynamics of the atomic Bose–Einstein condensate beyond the mean-field approximation",
Laser Phys. Lett. \textbf{18}, 055501 (2021).

\bibitem{Andreev JPP 21} P. A. Andreev,
"Hydrodynamics of quantum corrections to the Coulomb interaction via the third rank tensor
evolution equation: application to Langmuir waves and spin-electron acoustic waves",
J. Plasma Phys. \textbf{87}, 905870511 (2021).

\bibitem{Andreev Ch 21} P. A. Andreev,
"Quantum hydrodynamic theory of quantum fluctuations in dipolar Bose–-Einstein condensate",
Chaos \textbf{31}, 023120 (2021).

\bibitem{Andreev PoF 21} P. A. Andreev, I. N. Mosaki, and M. I. Trukhanova,
"Quantum hydrodynamics of the spinor Bose–Einstein condensate at non-zero temperatures",
Phys. Fluids \textbf{33}, 067108 (2021).


\bibitem{Andreev PRE 15} P. A. Andreev, "Separated spin-up and spin-down quantum hydrodynamics of degenerated electrons",
Phys. Rev. E \textbf{91}, 033111 (2015).

\bibitem{Andreev PP 16 SSE kin} P. A. Andreev,
"Spin-electron acoustic waves: The Landau damping and ion contribution in the
spectrum",
Phys. Plasmas \textbf{23}, 062103 (2016).


\bibitem{Ryan PRB 91} J. C. Ryan,
"Collective excitations in a spin-polarized quasi-two-dimensional electron gas",
Phys. Rev. B \textbf{43}, 4499 (1991).

\bibitem{Agarwal PRL 11} A. Agarwal, M. Polini, R. Fazio, G. Vignale,
"Persistent Spin Oscillations in a Spin-Orbit-Coupled Superconductor",
Phys. Rev. Lett. \textbf{107}, 077004 (2011).

\bibitem{Agarwal PRB 14} A. Agarwal, M. Polini, G. Vignale, M. E. Flatte,
"Long-lived spin plasmons in a spin-polarized two-dimensional electron gas",
Phys. Rev. B \textbf{90}, 155409 (2014).




\bibitem{Schober EPL 20} M. Schober, D. Kreil, and H. M. Bohm,
"Dynamic response of partially spin- and valley-polarised two-dimensional electron liquids",
EPL \textbf{129}, 17001 (2020).

\bibitem{Amico JPD 19} I. D'Amico, F. Perez, and C. A. Ullrich,
"Chirality and intrinsic dissipation of spin modes in two-dimensional electron liquids",
J. Phys. D: Appl. Phys. \textbf{52}, 203001 (2019).

\bibitem{Yang Ph 21} C. Yang, Z.-C. Wang, G. Su,
"Spin-polarized plasmon in ferromagnetic metals",
Physica A \textbf{575}, 126043 (2021).


\bibitem{Kreil CPP 18} D. Kreil, C. Staudinger, K. Astleithner, H. M. Bohm,
"Resonant and anti-resonant modes of the dilute, spin-inbalanced,
two-dimensional electron liquid including correlations",
Contrib. Plasma Phys. \textbf{58}, 179 (2018).


\bibitem{Kreil PRB 15} D. Kreil, R. Hobbiger, J. T. Drachta, and H. M. Bohm,
"Excitations in a spin-polarized two-dimensional electron gas",
Physical Review B \textbf{92}, 205426 (2015).


\bibitem{Huang PRB 16} C. Huang, Y. D. Chong, G. Vignale, and M. A. Cazalilla,
"Graphene electrodynamics in the presence of the extrinsic spin Hall effect", Physical Review B \textbf{93}, 165429 (2016)

\bibitem{Afanasiev PRB 22} A. N. Afanasiev,
"Acoustic plasmons and isotropic short-range interaction in two-component electron liquids",
Physical Review B \textbf{106}, 224301 (2022).


\bibitem{Andreev 2112} P. A. Andreev,
"Spin-electron-acoustic waves and solitons in high-density degenerate relativistic plasmas",
Phys. Plasmas \textbf{29}, 122102 (2022).
%


\bibitem{Andreev 2021 09} P. A. Andreev,
"Microscopic model for relativistic hydrodynamics of ideal plasmas",
arXiv:2109.14050.

\bibitem{Andreev 2021 11} P. A. Andreev,
"Relativistic hydrodynamic model with the average reverse gamma factor evolution for the degenerate plasmas:
high-density ion-acoustic solitons",
Phys. Plasmas \textbf{29}, 062109 (2022).




\bibitem{Landau v2} L. D. Landau and E. M. Lifshitz, \emph{The Classical Theory of Fields} (Butterworth-Heinemann, 1975).
L. D. Landau, E.M. Lifshitz, "Field theory" , Vol. 2 of Course of Theoretical Physics.



\bibitem{Andreev 2204} P. A. Andreev,
"Hydrodynamic and kinetic representation of the microscopic dynamics
as the transitions on the macroscopic scale of description and meaning of the self-consistent field approximation in these models",
arXiv:2204.11141.

\bibitem{Vlasov JETP 38} A. A. Vlasov,  J. Exp. Theor. Phys. \textbf{8},  291
(1938); A. A. Vlasov  Sov. Phys. Usp. \textbf{10}, 721 (1968).













%






\end{thebibliography}
\end{document}